\documentclass[12pt]{article}
\usepackage{amsmath,amssymb,exscale,color,axodraw,epsfig,multirow}
\input epsf
\definecolor{azul}{rgb}{0,0,1}
\definecolor{rojo}{rgb}{1,0,0}
\definecolor{verde}{cmyk}{0.64,0,0.95,0.50}
\definecolor{oliva}{cmyk}{0.64,0,0.95,0.40}
\definecolor{nara}{cmyk}{0,0.61,0.87,0}
\definecolor{rojonara}{cmyk}{0,0.77,0.87,0}
\definecolor{nararojo}{cmyk}{0,1,0.50,0}
\definecolor{marron}{cmyk}{0,0.87,0.68,0.32}
\definecolor{ladrillo}{cmyk}{0,0.89,0.94,0.28}
\definecolor{violerojo}{cmyk}{0,0.81,0,0}
\definecolor{rojoviole}{cmyk}{0.07,0.90,0,0.34}
\definecolor{purpura}{cmyk}{0.45,0.86,0,0}
\definecolor{violeta}{cmyk}{0.79,0.88,0,0}
\definecolor{azulviole}{cmyk}{0.86,0.91,0,0.04}
\definecolor{brown}{cmyk}{0,0.81,1,0.60}
\definecolor{negro}{cmyk}{0,0,0,1}
\definecolor{azulnoche}{cmyk}{0.98,0.13,0,0.43}
\definecolor{azulreal}{cmyk}{1,0.50,0,0}
\definecolor{azulpro}{cmyk}{0.96,0,0,0}
\definecolor{ama}{cmyk}{0,0,1,0}
\definecolor{amanara}{cmyk}{0,0.42,1,0}
\definecolor{amaver}{cmyk}{0.44,0,0.74,0}
\definecolor{verdeam}{cmyk}{0.15,0,0.69,0}
\definecolor{burntorange}{cmyk}{0,0.51,1,0}
\definecolor{cadetblue}     {cmyk}{0.62,0.57,0.23,0}
\definecolor{cornflowerblue}{cmyk}{0.65,0.13,0,0}
\definecolor{navyblue}      {cmyk}{0.94,0.54,0,0}
\definecolor{skyblue}       {cmyk}{0.62,0,0.12,0}
\definecolor{aquamarine}    {cmyk}{0.82,0,0.30,0}
\definecolor{gris}{cmyk}{0,0,0,0.50}
\definecolor{blanc}{cmyk}{0,0,0,0}

\newcommand{\m}[1]{\marginpar{{\tiny *}} }
\newcommand{\nova}[1]{}
\newcommand{\ttbar}{$t \bar t$}
\newcommand{\met}{{\not \!\! E_T}}

\begin{document}
\topmargin -1.0cm
\oddsidemargin -0.8cm
\evensidemargin -0.8cm

\thispagestyle{empty}
\vspace{20pt}

\hfill
\vspace{20pt}
\begin{center}
{\Large \bf Single production of an exotic bottom partner at LHC}
\end{center}

\vspace{15pt}
\begin{center}
{\large{Ezequiel \'Alvarez\footnote{sequi@df.uba.ar}$^{,a}$, Leandro Da Rold\footnote{daroldl@cab.cnea.gov.ar}$^{,b}$, Juan Ignacio Sanchez Vietto\footnote{jisanchez@df.uba.ar}$^{,a}$}}

\vspace{20pt}
$^{a}$\textit{CONICET, IFIBA and Departamento de F\'{\i}sica, FCEyN, Universidad de Buenos Aires \\
Ciudad Universitaria, Pab.1, (1428) Buenos Aires, Argentina}
\\[0.2cm]
$^{b}$\textit{CONICET, Centro At\'omico Bariloche and Instituto Balseiro\\
Av.\ Bustillo 9500, 8400, S.\ C.\ de Bariloche, Argentina}
\end{center}

\begin{abstract}
We study single production and detection at the LHC run II of exotic partners of the bottom quark. For masses larger than 1 TeV single production can dominate over pair production that is suppressed due to phase space. The presence of exotic partners of the bottom is motivated in models aiming to solve the $A_{FB}^b$ anomaly measured at LEP and SLC. Minimal models of this type with partial compositeness predict, as the lightest bottom partner, a new fermion $V$ of electric charge $-4/3$, also called {\it mirror}. The relevant coupling for our study is a $WVb$ vertex, which yields a signal that corresponds to a hard $W$, a hard $b$-jet and a forward light jet. We design a search strategy for the leptonic decay of the $W$, which avoids the large QCD multijet background and its large uncertainties. We find that the main backgrounds are $W+$jets and $t\bar t$, and the key variables to enhance the signal over them are a hard $b$-jet and the rapidity of the light jet.  We determine the discovery reach for the LHC run II, in particular we predict that, for couplings of order $\sim g/10$, this signal could be detected at a 95\% confidence level with a mass up to $2.4$ TeV using the first 100 fb$^{-1}$.
\end{abstract}

\newpage
\setcounter{footnote}{0}
\section{Introduction}

With the discovery of what seems to be the Higgs boson with a mass $m_h \approx 126$ GeV the ATLAS\cite{Aad:2012tfa} and CMS\cite{Chatrchyan:2012ufa} collaborations have initiated a long way in the understanding of the mechanism of Electroweak Symmetry Breaking (EWSB). For many years, this task has been organized around the naturalness principle which has been the starting point in the construction or proposal of most of the EWSB models. Two broad branches of study of EWSB theories can be identified: weakly coupled and strongly coupled ones. The main exponent of the former is Supersymmetry (SUSY) whereas for the latter one of the most interesting is Composite Higgs Models (CHM). 

A general feature of both of these Beyond the Standard Model (BSM) theories is the prediction of anomalous coupling of the Higgs Boson to the other particles of the Standard Model (SM). However, the 7/8 TeV LHC run I has shown no significant deviation from the SM expectation and results on the couplings of this Higgs-like particle to the other SM particles seem to prefer the original Higgs mechanism~\cite{Aad:2013wqa,Chatrchyan:2013lba}. Unfortunately, the LHC is not suitable to Higgs couplings precision measurements and in this respect any clue that could shed light into the mechanism of EWSB is very difficult to achieve~\cite{Peskin:2012we}. 

On the other hand, other common prediction in most of these models of EWSB is the presence of partners of the quarks of the third generation which are lighter than the other new particles~\cite{lightsusy,light}. In particular, since the top coupling with the Higgs is order 1, the top quark partners need to be light to stabilize the Higgs potential and keep naturalness. For LHC, searches for partners of the third generation of quarks is one of the best ways to pursue hints of new physics related to EWSB. In SUSY the top partners are spin 0 particles called {\it stops}. After the LHC run I no evidence for stops has been found up to 650 GeV putting some tension with natural SUSY~\cite{Craig:2013cxa}. In CHM the top partners are spin 1/2 vector-like fermions. ATLAS and CMS studies have constrained the top partners to be above 600-700 GeV at a 95\% confidence level~\cite{limits}. 

Besides the requirements to stabilize the Higgs potential, there are two ingredients of many CHM that we want to stress because they are responsible for important properties of these theories, with a deep impact in the phenomenology. First, to avoid large corrections to the oblique parameters in these theories, it is usual to consider that the composite sector has a global symmetry larger than the SM gauge one, containing the custodial symmetry~\cite{Agashe:2003zs}.~\footnote{See Ref.~\cite{Cabrer:2010si} for a different approach.} The composite resonances furnish complete representations of the extended symmetry of the composite sector. The choice of the representations of the composite fermions under this group defines different alternatives for the top partners. Second, the paradigm of partial compositeness: the masses of the SM fermions arise from mixing with composite fermions, that in turn couple with the Higgs~\cite{Kaplan:1991dc,Contino:2006nn}. Partial compositeness gives an economical mechanism to naturally obtain the hierarchy in the fermionic spectrum of the SM (although flavor mixing require some extra ingredients), it also gives rise to a mild separation of scales in the composite sector. Within partial compositeness, the large top mass requires large mixing, simultaneously leading to some top partners with masses parametrically smaller than the composite scale, these are the components of the multiplet that do not mix with the top before EWSB and are usually called {\it custodians}~\cite{light,Pomarol:2008bh}. These states, being usually the lightest new particles, are the leading candidates to direct searches at colliders. The discovery signatures of top-like heavy quarks (${T}$) at LHC have been studied in the context of CHM's, for instance, in Refs.~\cite{Carena:2007tn,single}. Bottom-like ($B$) and exotic quark of charge 5/3 ($X$) have been widely discussed also~\cite{samesign}. These studies have motivated many searches that have putted stringent limits on the masses of the top partners. These searches assume model-independent QCD pair production and the model-dependent part only modifies the weight of each decay channel. Recently, the authors of Ref.~\cite{DeSimone:2012fs} have reassessed these limits for the case of Pseudo Goldstone Boson Higgs, showing that in some cases one can exclude top partners with masses up to 1.5 TeV. For such masses, single production of exotic quarks, although usually being electroweak (EW) suppressed, starts being competitive with double production~\cite{Willenbrock:1986cr}. Almost all these searches have been restricted to top partners, assuming that the bottom partners are  heavier.~\footnote{However Refs.~\cite{AguilarSaavedra:2009es,Aguilar-Saavedra:2013qpa} have also considered a extended list of partners of third generation quarks in different representations of SU(2)$_L\times$U(1)$_Y$.}

In this work we explore another possibility motivated by the third generation anomalies in LEP and SLC. One of the largest known tension of a light Higgs with the data is in the bottom forward-backward asymmetry ($A^b_{FB}$) at the Z pole in LEP and SLC. For a light Higgs, the deviation in this observable is about 2.9$\sigma$ compared with the best global fit, suggesting a modification of the $Zb\bar b$ coupling. On the other hand, the branching ratio of $Z$ decaying to a pair bottom anti-bottom ($R_b$) is in very good agreement with the SM expectation~\cite{ALEPH:2010aa}. Explaining the $A^b_{FB}$ deviation without simultaneously spoiling the agreement for $R_b$ requires in general extra structure~\cite{Choudhury:2001hs}. Within the framework of CHM, the large shift of $Zb_R\bar b_R$ coupling needed to solve this puzzle requires large mixing of $b_R$, then partial compositeness leads to light bottom partners with masses parametrically smaller than the composite scale. In this work we study the possibility to produce and detect the bottom partners at the LHC, within an effective CHM that addresses the bottom puzzle and can accommodate the mass spectrum of the third generation of quarks~\cite{DaRold:2010as}, see also~\cite{Djouadi:2011aj}. We will consider a minimal realization in terms of a two-site model that allows to compute the couplings and the spectrum of resonances.~\footnote{In Ref.~\cite{Alvarez:2010js} the model was extended to explain the value of the top forward-backward asymmetry ($A^t_{FB}$) measured by CDF and D0 collaboration at Tevatron~\cite{Aaltonen:2012it,Abazov:2011rq}, that also have shown a considerable deviation from the SM expectation.} 
The model can be extended to include the Higgs as a Pseudo Goldstone Boson. 

In the minimal CHM with custodial symmetry the bottom composite partners include bottom and top-like resonances, as well as exotic resonances of charge -4/3 ($V$)~\footnote{Other common quotation for an exotic particle or mirror of charge -4/3 is $\chi$~\cite{Choudhury:2001hs,Kumar:2010vx} and $Y$~\cite{AguilarSaavedra:2009es,Aguilar-Saavedra:2013qpa}.}  and -7/3 ($S$), also called mirrors.
Due to the large mixing of $b_R$, the lightest of these resonances is a custodian $V$, a partner of $b_R$. $V$ can be produced through QCD pair production or EW single production. Having charge -4/3, and assuming suppressed mixing with light generations, it can be single produced only through the vertex $WVb$ with just one decay channel for the exotic fermion: $V \to bW^-$. Thus, as long as the bottom charge is not measured, the signature for pair and single production is analogue to $T$ if its decays are exclusively through the $bW$ channel or to a top-like quark of a chiral fourth generation. Therefore, the limit for pair production of $T$ when it decay exclusively through $T \to bW$, applies, being this $m_{V} > 740 $ GeV~ at a 95\% confidence level \cite{ATLAS_T}. 

The present 5$\sigma$ discovery reach for $V$ through QCD pair production is estimated to be 820 GeV for the early 14 TeV LHC run II of 100 fb$^{-1}$~\cite{AguilarSaavedra:2009es}. As the discovery reach of the LHC for heavy quarks approaches masses around TeV scale, pair production begins to loss power of discovery against single production due to phase space suppression. The goal of this work is to design a search strategy that works in the range of masses where the EW single production of heavy quarks dominates over pair production. This search strategy is suitable for both: $V$ and $T$, whether the later decays exclusively to $bW$, though our study is motivated by the former.  A model of $T$ being the lightest new resonance of the New Physics sector and being produced and decaying predominantly through $VbW$ vertex is hard to justify, whereas those properties are guaranteed for $V$ in the model of Ref.~\cite{DaRold:2010as}.

As was pointed out in Ref.~\cite{DeSimone:2012fs} present experimental searches are not sensitive to single production of third generation partners. This is the case with searches for single production of a new quark which decays to one $b$-jet and a $W$. Although there are many experimental searches for sequential fourth generation of quarks with this signature, they usually assume $b'$ and $t'$ to be close in mass and lighter than 1 TeV, as required by EW precision tests and perturbativity. Thus, these searches are inclusive on both pair and single production of either $t'$ and/or $b'$. However, in these conditions channels with more multiplicity in $W$ and/or $b$-jets are more relevant, and these are the ones the experimental studies have been concentrated in so far.

Single production jointly with pair production of bottom partners was first studied in Ref.~\cite{Kumar:2010vx}, where the authors considered a single value for the coupling and showed their results for a mass of the exotic fermion of 500 GeV. Ref.~\cite{Aguilar-Saavedra:2013qpa} has also analyzed the allowed single production cross-sections at the LHC. The region of the parameter space and the search strategy that we propose in this work is different from those studies. We will propose a search strategy that relies on main features of the EW single production of $V$ as a high $p_T$ $b$-jet and a forward light jet. Previous works in this respect can be found in Refs.~\cite{single}. We will propose here a new channel with only one tagged $b$-jet, one lepton, missing energy and a forward light jet that improves the sensitivity of the early LHC run II. Contrary to QCD pair production, EW single production of resonances depends not only on the mass of the resonance, but also on the EW coupling with SM fields. Thus in a reduced picture, the relevant parameters are the mass and EW coupling of the resonances. One of the main results of our work is to determine the region in this parameter space where LHC can discover the bottom partner. We find that our complementary search strategy extends the 5$\sigma$ (2$\sigma$) discovery reach for $V$ from 820 GeV to roughly 1.7 TeV (2.4 TeV) for couplings ${\cal O}(g/10)$. Even so, in the intermediate region 800-1000 GeV an enhancement on the sensitivity can be achieved taking advantage of both, QCD pair and EW single production together. We leave this for future work.

The structure of this work is as follows: In section 2 we briefly describe the effective model, we show the embedding of the top and bottom partners fields into the global symmetry of the new strongly coupled sector and describe the spectrum of the mass eigenstates and its couplings with the SM. In section 3 we discuss the production mechanism and decays of the $V$-resonance predicted by the model. In section 4 we describe the kinematical features of the signal and main backgrounds for the single production of $V$. Then, we design a cut-based search strategy for the signal. Finally we show the reach of the search strategy for the early 14 TeV LHC run II and expected limits for 300 and 500 fb$^{-1}$. We end with some discussion and conclusions in sections 5 and 6.

\section{The model}
We give a brief description of the model of Ref.~\cite{DaRold:2010as}, where effects from a new strongly interacting sector can solve the $A^b_{FB}$ anomaly of LEP and SLC by shifting the $Zb\bar b$ couplings. We will focus on the spectrum of bottom partners and their couplings to the SM fields relevant for single creation of resonances at LHC. For more details on this kind of effective theories we refer the reader to the original reference and to~\cite{Contino:2006nn}.

We consider a model with two sectors, an elementary one, whose field content is as in the SM except for the Higgs, and a new sector with strong interactions that lead to resonances with masses $m_{\phi^{cp}}$ of order TeV, plus a lighter scalar boson corresponding to the Higgs field. We will assume that the interactions between the resonances can be described by couplings $g_{cp}$ involving vector resonances and $y_{cp}$ corresponding to proto Yukawa interactions, such that: $g_{SM}\ll g_{cp}\ll 4\pi$ and $y_{cp}\sim 1-2\pi$.

In the minimal set-up with custodial symmetry the composite sector has a global symmetry [SU(3)$_c\times$SU(2)$_L\times$SU(2)$_R\times$U(1)$_X]^{cp}$, with vector resonances in the adjoint representation, and hypercharge realized as $Y=T^{3R}+X$. The leading order interactions involving the vector resonances can be obtained by use of covariant derivative on the composite sector. The Higgs field $\Sigma=(\tilde H,H)$ is a bidoublet of the composite symmetry $({\bf 2},{\bf 2})_0$ and it does not couple to U(1)$_X$. A Pseudo Goldstone Boson realization of the Higgs with dynamical EW symmetry breaking can be obtained by extending the EW symmetry of the strong sector to SO(5)$\times$U(1)$_X$ with a spontaneous breaking of SO(5) to SO(4)~\cite{Agashe:2004rs}. The strong dynamics of the composite sector also leads to fermionic resonances.

The elementary fermions mix linearly with operators of the strongly coupled sector realizing the idea of partial compositeness
\begin{equation}
{\cal L}\supset y_L\bar \psi^{el}_L {\cal P}_\psi{\cal O}_R + y_R\bar {\tilde\psi}^{el}_R {\cal P}_{\tilde\psi}\tilde {\cal O}_L + {\rm h.c.} \ ,
\end{equation}
with ${\cal O}_R$ and $\tilde {\cal O}_L$ fermionic operators of the strong sector. Since the symmetry of the composite sector is larger than the SM one, we have introduced projectors ${\cal P}_\psi$ that project the composite multiplets onto the components with the quantum numbers of the SM fields. Assuming that the composite operators can create fermionic resonances with masses of order TeV, at low energies partial compositeness aims to linear mixing with them
\begin{equation}
{\cal L}\supset \bar \psi^{el}_L \Delta{\cal P}_{\psi}\psi^{cp}_R + \bar {\tilde\psi}^{el}_R \tilde\Delta{\cal P}_{\tilde\psi}\tilde\psi^{cp}_L + {\rm h.c.} \ ,
\end{equation}
where $\Delta$ and $\tilde\Delta$ have mass dimension and parameterize the mixing. As we will show explicitly below, partial compositeness allows to obtain small masses for the light SM fermions by taking the mixing of at least one of the chiralities to be small compared with the composite scale. 

We will consider just the third generation, assuming that both chiralities of the light quarks have small mixing and can be neglected in our analysis. If the structure of the strong sector is rich enough, it is possible to mix the elementary fermions with several operators in different representations of the composite group, each operator having its own coupling. In fact, as explained in Ref.~\cite{DaRold:2010as}, to solve the bottom puzzle and obtain the masses of the third generation quarks, a model with two resonances mixing with $q_L^{el}$ is preferred, ${\cal L}\supset \bar q^{el}_L (\Delta_1{\cal P}_1q^{cp}_{1R}+\Delta_2{\cal P}_2q^{cp}_{2R})$. Below we specify the quantum numbers of these resonances.

Similar to the fermions, there is mixing between the SM gauge fields and the bosonic operators of spin one of the strong sector. At low energies it is enough to consider the mixing with the lightest level of vector resonances created by those operators, with TeV masses $m_{A^{cp}}$ arising from the strong dynamics. The mixing preserves the diagonal subgroup G$_{\rm el+cp}$, leading to a set of massless fields that correspond to the SM gauge symmetry. Matching at tree level with the couplings of the SM leads to $g_{SM}=g_{el}g_{cp}/\sqrt{g_{el}^2+g_{cp}^2}$.

Before EWSB there is a set of massless fermions $q_L$, $t_R$ and $b_R$, and gauge bosons $A_\mu$, with the same quantum numbers as in the SM. These states can be obtained by performing a rotation between the elementary and composite states~\cite{Contino:2006nn}
\begin{eqnarray}
\begin{bmatrix}\phi \\ \phi^*\end{bmatrix}=\begin{bmatrix}\cos\theta_\phi & \sin\theta_\phi \\ -\sin\theta_\phi & \cos\theta_\phi \end{bmatrix} \begin{bmatrix} \phi^{el} \\ {\cal P}_\phi \phi^{cp} \end{bmatrix} \ , 
\qquad \phi=A,\psi_L,\tilde\psi_R \ , \label{rot1} \\
\tan\theta_A=\frac{g_{el}}{g_{cp}} \ , 
\qquad \tan\theta_\psi=\frac{\Delta_\psi}{m_{\psi^{cp}}} \ , \qquad \tan\theta_{\tilde\psi}=\frac{\Delta_{\tilde\psi}}{m_{\tilde\psi ^{cp}}} \ ,
\label{rot2} 
\end{eqnarray}
with $A_\mu$, $\psi_L$ and $\tilde\psi_R$ the massless fields and $A^*$, $\psi^*$ and $\tilde \psi^*$ the combination of composite and elementary fields that remains massive, with mass $M_{\phi^*}=m_{\phi^{cp}}/\cos\theta_\phi$. The multiplets of resonances contain new states that do not mix with the elementary ones before EWSB, the custodians, that can be defined as $\tilde{\cal P}_\phi \phi^{cp}\equiv(1-{\cal P}_\phi)\phi^{cp}$. The mass of the custodians is suppressed compared with the other components of a multiplet: $M_{\tilde{\cal P}_\phi \phi}=m_{\phi^{cp}}$. In the rest of this work we will fix the scale $M_{\phi^*}=M$ to be the same for all the fields. The custodian mass depends on this scale and on the size of the mixing: $M_{\tilde{\cal P}_\phi \phi}=M\cos\theta_{\phi}$, thus for those fields with large mixing the mass of the corresponding custodians will be parametrically smaller than the composite scale.

After diagonalization of the elementary/composite mixing, the proto Yukawa interactions lead to interactions between the Higgs and the would be massless fermions:
\begin{equation}\label{mfermionSM}
{\cal L}\supset y_{cp}\ \sin\theta_\psi\ \sin\theta_{\tilde\psi}\ \bar\psi_L h\tilde\psi_R\ . 
\end{equation}
After EWSB these interactions are responsible for the mass of the SM fermions, that are controlled by the size of the mixing of each chiral fermion.

The quantum numbers of the bottom partners are chosen to induce tree level shifts in $Zb\bar b$ couplings that can accommodate the experimental results on $A^b_{FB}$ and $R_b$. In the minimal model we consider resonances $q_2^{cp}$ and $b^{cp}$ mixing respectively with $q^{\rm el}_L$ and $b^{\rm el}_R$
\begin{eqnarray}
q_2^{cp}=({\bf 2},{\bf 3})_{-5/6}=
\begin{bmatrix}
 V''^{cp}_2 & D'^{cp}_2 & U^{cp}_2 \\
 S'^{cp}_2 & V'^{cp}_2 & D^{cp}_2 
\end{bmatrix} \ , 
&  b^{cp}=({\bf 1},{\bf 2})_{-5/6}=\begin{bmatrix} V'^{cp}_b & D^{cp}_b\end{bmatrix} \ , \label{bembeddingb}
\end{eqnarray}
where $({\bf r}_L,{\bf r}_R)_{{\bf r}_X}$ denotes the representation for [SU(2)$_L\times$SU(2)$_R\times$U(1)$_X]^{cp}$.~\footnote{It is possible to consider larger representations also~\cite{DaRold:2010as}.} $V$ and $S$ are exotic fermions with $Q=-4/3$ and $-7/3$, respectively, primed fermions are custodians. The mixing term with the elementary fermions explicitly breaks SU(2)$_R$, and requires projectors ${\cal P}_{b_L}$ and ${\cal P}_{b_R}$ that select the proper components of the multiplets: ${\cal P}_{b_L}q_2^{cp}=(U_2^{cp},D_2^{cp})^t$ and ${\cal P}_{b_R}b^{cp}=D_b^{cp}$. 
As in Eq.~(\ref{mfermionSM}), the proto Yukawa $y_{cp}^b\bar q^{cp}_2\Sigma b^{cp}$ leads, after mixing, to the bottom mass. The size of the corrections of the couplings as well as the bottom mass are controlled by the mixing angles. 

Since the set of resonances in $q^{cp}_2$ and $b^{cp}$ do not allow the generation of the top mass, extra resonances are needed. We add to our model two new resonances, $q^{cp}_1$ and $t^{cp}$, mixing respectively with $q^{el}_L$ and $t^{el}_R$, the top mass arising from a composite proto Yukawa $y_{cp}^t\bar q_1^{cp}\Sigma t^{cp}$. The large top mass requires large mixing with both chiralities of the top, inducing dangerous corrections to $Zb_L\bar b_L$ interactions. To protect $g_{b_L}$ one can invoke a $P_{LR}$ symmetry for these resonances~\cite{Agashe:2006at}, demanding that $q_1^{cp}$ transforms as $({\bf 2},{\bf 2})_{2/3}$. Invariance of the proto Yukawa interaction under the composite symmetry requires $t^{cp}$ to be a $({\bf 1},{\bf 1})_{2/3}$ or a $({\bf 3},{\bf 1})_{2/3}+({\bf 1},{\bf 3})_{2/3}$. We choose the smallest representation for this work:
\begin{eqnarray}
q_1^{cp}=({\bf 2},{\bf 2})_{2/3}=
\begin{bmatrix}
 U^{cp}_1 & X'^{cp}_1 \\
 D^{cp}_1 & U'^{cp}_1 
\end{bmatrix} \ , 
&  t^{cp}=({\bf 1},{\bf 1})_{2/3}=U^{cp}_t \ , 
\label{bembeddingt}
\end{eqnarray}
with $X_1'^{cp}$ an exotic resonance with charge $5/3$, again primed fermions are custodians. $q_L^{el}$ and $t^{el}_R$ mix with ${\cal P}_{t_L}q_1^{cp}=(U_1^{cp},D_1^{cp})^t$ and $U^{cp}_t$, respectively. In this minimal embedding, the large mixing $\Delta_1$ leads to light $X_1'$ and $U_1'$ that could be produced at the LHC~\cite{Carena:2007tn,samesign}.

To study the production and decay of the bottom partners we need the spectrum and couplings in the mass basis. In the following we will argue which are the lightest states and we will estimate the size of their EW couplings, however we have checked that our estimates agree with the full numerical diagonalization. In fact for our scan we have made several considerations: we have matched the SM gauge couplings as previously explained and we have varied $1/8\leq g_{el}/g_{cp}\leq 1/5$, we have selected points of the parameter space that solve the bottom puzzle and reproduce the SM spectrum of the third generation and gauge bosons, we have considered natural Yukawa couplings $1/3\leq y_{cp}\leq 2\pi$, we have considered a composite scale $M\sim 2-3$ TeV. With these constraints we have checked that it is possible to obtain masses and couplings for the lightest resonance as those shown in the simulations. The mass matrices necessary for the calculations of the physical masses and couplings are shown in Appendix~\ref{ap_diag} in the gauge basis. 

Let us first analyze the spectrum of fermions. After EWSB all the fermions with equal charge are mixed. We will order the heavy fermions in the eigenmass basis according to increasing masses, {\it ex}: there are three exotic states with charge $-4/3$: $\{V'^{cp}_2,V''^{cp}_2,V'^{cp}_b\}$, whereas the mass basis will be $\{V_1,V_2,V_3\}$, with $m_{V_1}\leq m_{V_2}\leq m_{V_3}$. To gain some insight we consider first the situation of no EWSB, in this case there are no mixing between the $V$-states and the mass basis coincides with the gauge basis, the masses depending on the size of the elementary/composite mixing angles. The large $\delta g_{b_R}$ needed to solve the $A^b_{FB}$ anomaly suggests that $\theta_b$ should be larger than $\theta_2$, leading to $V_1= V'^{cp}_b$ before EWSB. In fact, in this case $V_b'^{cp}$ is the lightest bottom partner, providing the motivation for the study of $V$ production at the LHC. Varying $\theta_b$ we can obtain moderate to large suppression of $m_{V_1}$. If $\theta_b$ is large, the small ratio $m_b/m_t$ requires small $\theta_2$ and the masses of $V'^{cp}_2$ and $V''^{cp}_2$ are $\simeq M_{cp}+O(\theta_2^2)$, {\it ie}: their mass is approximately given by the composite scale. 
After EWSB there is mixing between the $V$-resonances induced by the Higgs, with strength $y_{cp}^b$. For moderate values of the Yukawa coupling, $y_{cp}^b\sim 1$, that are favored by the bottom mass, we have checked that for sizable $\theta_b$ the lightest state has a dominant projection on $V'^{cp}_b$. 

According to the arguments of the previous paragraph, the other custodians arising from the bottom sector are expected to be heavier. However, it is important to consider also the mass of the $B$-resonances in some detail, because in the case of $m_{B_1}< m_W+m_{V_1}$, $V_1$ can decay to $WB_1$ with strength ${\cal O}(y_{cp}^b)$, that can dominate over $Wb$. $B_1$ would decay to $hb$, $Zb$ and $Wt$, leading to a very different signal for $V$-production. The only $B$-custodian of the simplest model is $D'^{cp}_2$, whose mass is suppressed by $\cos\theta_2\sim1+{\cal O}(\theta_2^2)$, similar to $V'^{cp}_2$ and $V''^{cp}_2$. Therefore, in the case we are interested in, with $\theta_b\gg\theta_2$, $B$-resonances are always heavier than $V_1$, that decays exclusively by $V_1\to Wb$.

\begin{figure}
\begin{center}
\begin{picture}(250,110)
        \DashLine(0,50)(50,50){4}
	\Vertex(50,50){2}
	\ArrowLine(80,80)(50,50)
	\ArrowLine(50,50)(80,20)
	\Vertex(80,20){2}
	\ArrowLine(80,20)(110,20)
        \Text(0,60)[]{$h^-$}
        \Text(65,75)[]{$\bar V'^{cp}_b$}
        \Text(55,25)[]{$D^{cp}_2$}
        \Text(110,30)[]{$b^{el}_L$}
        \Text(65,50)[]{$y_b$}
        \Text(80,10)[]{$\Delta_2$}
        \DashLine(150,50)(200,50){4}
	\Vertex(200,50){2}
	\ArrowLine(230,80)(200,50)
	\ArrowLine(200,50)(230,20)
	\Vertex(230,20){2}
	\ArrowLine(230,20)(260,20)
        \Text(150,60)[]{$h^-$}
        \Text(215,75)[]{$\bar V'^{cp}_2$}
        \Text(205,25)[]{$D^{cp}_b$}
        \Text(260,30)[]{$b^{el}_R$}
        \Text(215,50)[]{$y_b$}
        \Text(230,10)[]{$\Delta_b$}
%
\end{picture}
\caption{Charged Higgs interactions involving V-resonances and bottom quark.}
\label{fig_ET}
\end{center}
\end{figure}
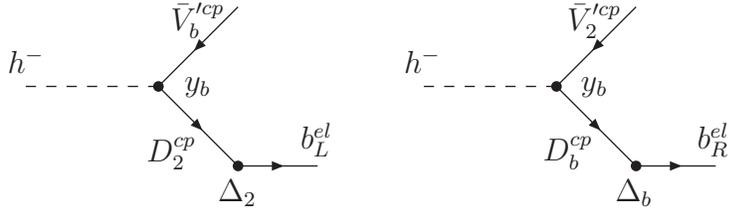

Single production of $V$ is driven by the interaction $WVb$. The size of this interaction can be computed by diagonalizing all the mixing. To obtain an estimate of its strength and its parametric dependence, it is simpler to consider the interaction with longitudinal EW gauge bosons only, that can be computed using the Equivalence Theorem. We have to consider the charged Yukawa interaction $h^-\bar Vb+{\rm h.c.}$ that in the gauge basis arises from the proto Yukawa between composite fermions and the Higgs:
\begin{equation}\label{b-yukawa}
{\cal L}\supset y_{cp}^b \bar q^{cp}_2 \Sigma b^{cp} +{\rm h.c.}\supset y_{cp}^b \left[h^-\left(\frac{1}{\sqrt{2}}\bar V'^{cp}_b D^{cp}_2+\bar D^{cp}_b U^{cp}_2\right)+h^+\left(\frac{1}{\sqrt{2}}\bar D^{cp}_b V'^{cp}_2+\bar V'^{cp}_b S'^{cp}_2\right)\right] \ .
\end{equation}
After diagonalization of the elementary/composite mixing Eq.~(\ref{b-yukawa}) leads to interactions involving the bottom quark. In Fig.~\ref{fig_ET} we show the leading order interaction with the bottom quark expanding in powers of elementary/composite mixing insertions. For the lightest resonance $V_1\simeq V^{cp'}_b$, to leading order in elementary/composite mixing the interaction $hVb$ can be approximated by $g_{1} h^-\bar V_{1R} b_L+{\rm h.c.}$ with $g_{1}\simeq y_{cp}^b\sin\theta_2\sim O(10^{-1})$, whereas for $V_2\simeq V^{cp'}_2$ the leading interaction is $\tilde g_{2} h^-\bar V_{2L} b_R$ with $\tilde g_{2}\simeq y_{cp}^b\sin\theta_b\sim O(1)$. Thus, for $\theta_b\gg\theta_2$, the lightest resonance has a smaller coupling with $W_L$, decaying preferentially with $R$-polarization, whereas the next $V$-resonance has a larger coupling with $W_L$ and decays preferentially with $L$-polarization.

After EWSB there are corrections to the estimates made before, but the order of magnitude does not change. We have verified numerically that the order of magnitude of the mixing required to solve the bottom puzzle are as in the previous paragraph (see Ref.~\cite{DaRold:2010as} for analytic estimates), resulting in $g_{1}\lesssim 0.1$ and $\tilde g_{2}\lesssim 1$.

From now on we will consider just the lightest $V$-resonance, we will denote its mass $m_V$, and its couplings with $Wb$ as $g_{L,R}$. Moreover, since $g_R\gg g_L$, we will consider the effect of $g_R$ only, neglecting $g_L$. We will restrict $g_R < 0.065$, since larger values are hard to be obtained satisfying the conditions explained before.

\section{Single production and decay of $V$}
The model described in the previous section predicts many new particles and, therefore, many possible signatures which could be detected at the LHC.  Although the model has a range where the masses and effective couplings could vary, the embedding of the fermions predicts that the lightest particle should be an exotic quark $V$.  We study in this section the production mechanisms for $V$ mirror quark and its dependence on the parameters of the model.

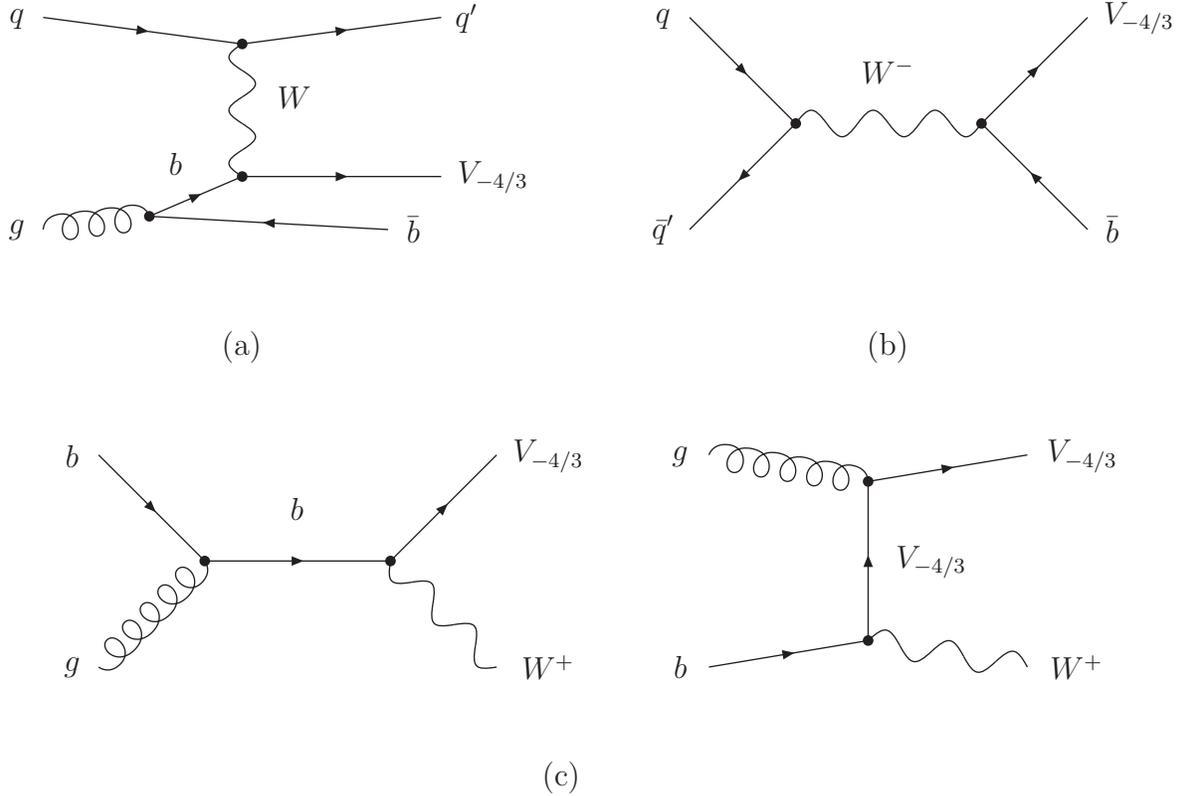
\begin{figure}
\begin{minipage}[b]{0.5\textwidth}        
\begin{center}
\begin{picture}(150,100)	
  	\ArrowLine(0,100)(75,90) 		
	\ArrowLine(75,90)(150,100) 		
 	\Gluon(0,20)(40,25){5}{3}
	\ArrowLine(40,25)(75,40)			
	\ArrowLine(130,20)(40,25)		
	\ArrowLine(75,40)(150,40)		
	\Photon(75,40)(75,90){5}{2.5}		
	
	\Vertex(75,90){2}
	\Vertex(40,25){2}
	\Vertex(75,40){2}

   	\Text(-10,100)[]{$q$}			 		
	\Text(-10,20)[]{$g$}			 			
	\Text(160,100)[]{$q'$}					
	\Text(140,20)[]{$\bar b$}
	\Text(170,40)[]{$V_{-4/3}$}
	\Text(50,45)[]{$b$}
	\Text(95,70)[]{$W$}
\end{picture}
\end{center}
\center{(a)}	
\end{minipage}
\begin{minipage}[b]{0.5\textwidth}
\begin{center}
\begin{picture}(150,100)	
    \ArrowLine(0,100)(40,60) 		
	\ArrowLine(40,60)(0,20) 			
 	\Photon(40,60)(110,60){5}{3}		
	\ArrowLine(110,60)(150,100)		
	\ArrowLine(150,20)(110,60)		
	
	\Vertex(40,60){2}
	\Vertex(110,60){2}
	
   	\Text(-10,100)[]{$q$}			 		
	\Text(-10,20)[]{$\bar q'$}			 	
	\Text(170,100)[]{$V_{-4/3}$}				
	\Text(160,20)[]{$\bar b$}
	\Text(75,80)[]{$W^-$}	
\end{picture}
\end{center}
\center{(b)}
\end{minipage}
\vskip 1.2cm     
\begin{center}    
\begin{picture}(350,100)	
    \ArrowLine(0,100)(40,60) 			
	\Gluon(40,60)(0,20){5}{5}	
 	\ArrowLine(40,60)(110,60)			
	\ArrowLine(110,60)(150,100)			
	\Photon(150,20)(110,60){5}{2.5}		
	
	\Vertex(40,60){2}
	\Vertex(110,60){2}
	
   	\Text(-10,100)[]{$b$}			 		
	\Text(-10,20)[]{$g$}					 	
	\Text(170,100)[]{$V_{-4/3}$}				
	\Text(170,20)[]{$W^+$}
	\Text(75,80)[]{$b$}	

	\Gluon(230,100)(290,90){5}{5} 		
	\ArrowLine(290,90)(350,100)	 		
 	\ArrowLine(230,20)(290,30)			
	\ArrowLine(290,30)(290,90)			
	\Photon(290,30)(350,20){5}{2.5}		
	
	\Vertex(290,90){2}
	\Vertex(290,30){2}
	
   	\Text(220,100)[]{$g$}			 		
	\Text(220,20)[]{$b$}			 			
	\Text(372,100)[]{$V_{-4/3}$}				
	\Text(370,20)[]{$W^+$}	
	\Text(315,60)[]{$V_{-4/3}$}
\end{picture}
\center{(c)}
\end{center}
\caption{Single-$V$ production diagrams:  (a) $t$-channel, (b) $s$-channel and (c) $WV$ associated production. The $t$-channel diagram dominates because has the smallest suppression from the propagator.}
\label{production}
\end{figure}

\begin{figure}[!htb]
\begin{center}
\includegraphics[width=0.45\textwidth]{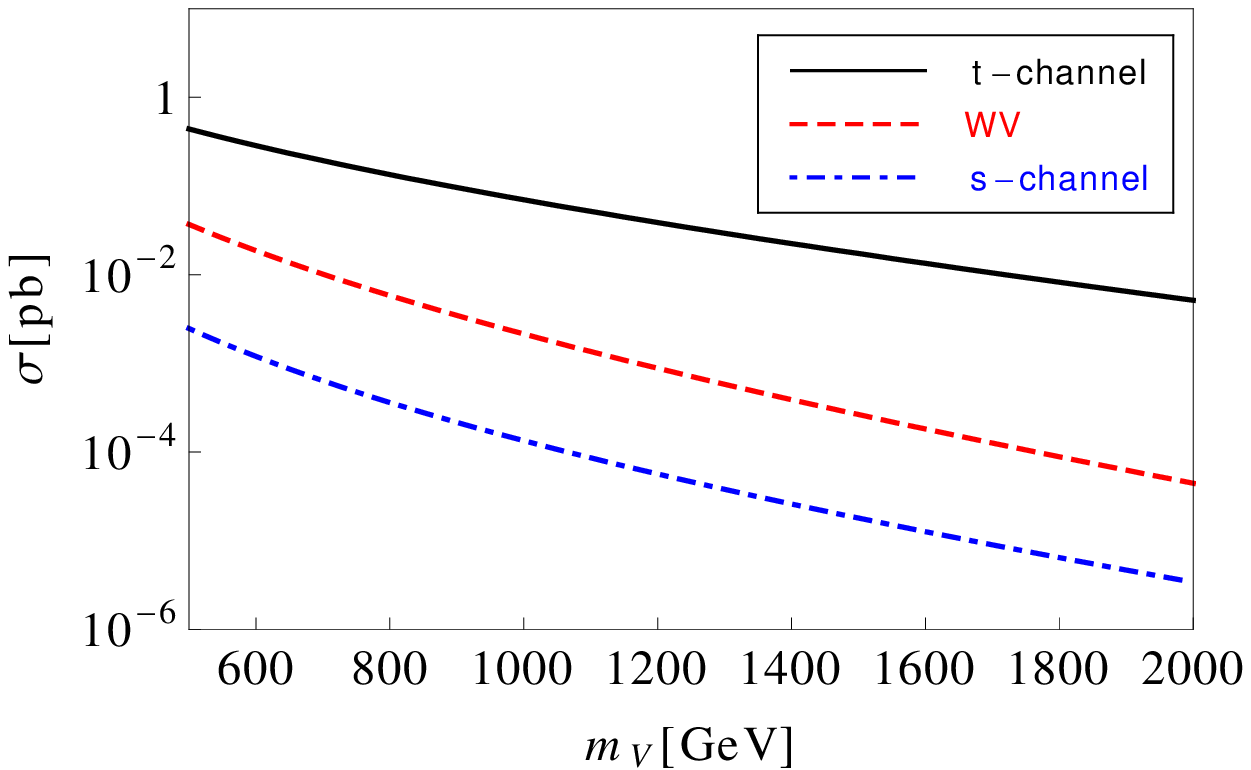}
~
\includegraphics[width=0.45\textwidth]{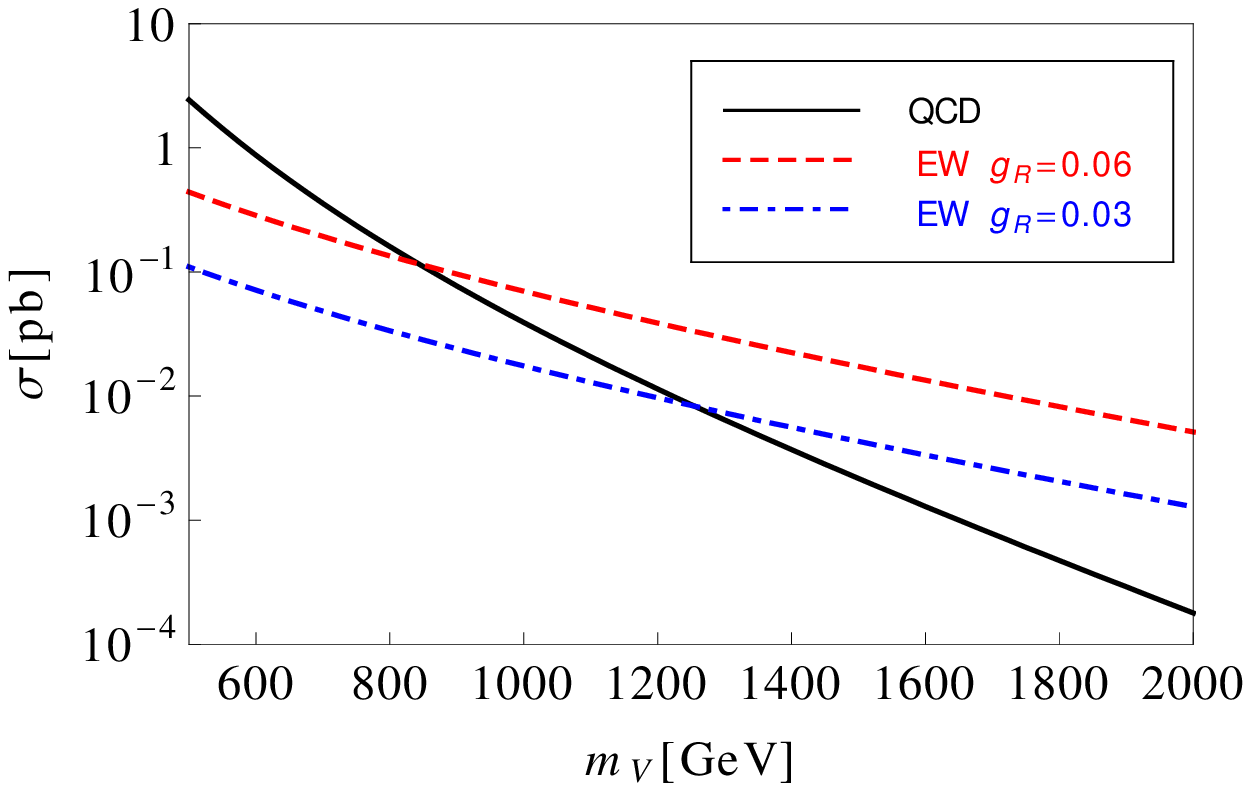}
\end{center}
\caption{[color online] Left panel: single-$V$ production cross-sections for the different production mechanisms with $g_R = 0.06$. Right panel: single- versus pair-$V$ production cross-section as a function of its mass for different $WVb$ couplings. Since in the model $g_R \gg g_L$, in both figures we set $g_L = 0$.}
\label{xsections}
\end{figure}

Being $V$ a colored particle, its coupling to gluons is model independent and $V$-pair production dominates in the low mass regime.  As the $V$ mass reaches 1 TeV and beyond, the phase space at the LHC at 14 TeV  suppresses pair production and single $V$ production has to be taken into consideration. Single production, however, depends on the coupling in the $WVb$ term which is model dependent. Since the $V$ interaction with $Wb$ is similar to the top quark in the SM, then it is natural to expect similar diagrams as in single-top production. In fact, single $V$ production can go through a $t$-channel, a $s$-channel or in associated production with a $W$ (see Fig.~\ref{production}).  As it can be expected, the $t$-channel dominates over the other possibilities.  The reason for this is that the $t$-channel amplitude has the smallest suppression from the propagator.  
(Similar reasoning holds for single-top production.)  We have plotted the cross-section for the different single-$V$ production mechanisms as well as for $V$-pair production as a function of the mass and the coupling in Fig.~\ref{xsections}.  As it can be seen from the figure, we can expect at least an order of magnitude more single- than pair- $V$ production for couplings $\sim 0.065$ and masses above $\sim 1.4$ TeV.

Once a single $V$ has been produced, its decay will go through the same vertex to a $W$ and a $b$ quark, which we assume to be the only decay channel for $V$. Typical widths for the $V$ are in the order of 1-100 GeV for the 1.3-2.5 TeV mass range and couplings in the $0.01-0.06$ range. These widths corresponds to a maximum lifetime of 10$^{-24}$ seconds which yields a vertex displacement of 3 $\times$ 10$^{-16}$ meters for the typical LHC run II collisions.  

In the case the $W$ decays leptonically, and assuming $t$-channel production, the process will consist in a hard $b$-jet, a hard lepton, $\not \hskip-0.1cm E$ from the neutrino, a forward light-jet, and a forward $b$-jet that comes from the gluon splitting in the proton,
$$ pp \to V\,b\,j \to W\,b\,b\,j \to \ell\,\not \hskip-0.1cm E\,b\,b\,j .$$
For a hadronic decaying $W$ the signature would be $b\,b\,j\,j\,j$, which would have an irreducible QCD multijet background. Since the QCD multijet simulation has large uncertainties, controlling this background requires data-driven methods which are beyond the scope of this work. For this reason we choose to work with leptonically decaying $W$ at the price of reducing in about a third the signal cross-section times branching ratio. However, groups with access to control samples should consider the hadronic channel also.

\section{Analysis and search strategy for V-single production}

\subsection{Signal features and Backgrounds}

The EW single production of the V quark depends only on the V mass and the left and right couplings to W. Since $g_R\gg g_L$, we are neglecting effects of $g_L$ and the cross-section scales with the couplings as $g_R^2$. This is true also for the QCD NLO correction of the cross-section. This allow us to use the results of the Ref.~\cite{Campbell:2009gj} where the NLO K-factor of the single production of heavy quarks between 1.3 TeV and 2.5 TeV range over 1.26 - 1.49. This leads to a cross-section of 1-10 fb which yields 100-1000 of single production events in the first 100 fb$^{-1}$ of the 14 TeV LHC run II. 

Being V a very heavy particle it is expected to be produced with low $p_T$. For the same reason, the $W$ and $b$ from its decay will be boosted and therefore approximately back-to-back in the laboratory. Therefore, the signal features a high $p_T(b)$ and $p_T(W)$. Hence, we expect the $b$-jet to be the highest $p_T$ particle and the lepton and missing energy from the decay of $W$ having high $p_T$. Observe that being the single production a $t$-channel process the jet that comes with the V is likely to be produced forward. Notice that this is true even with $W$-exchange because it is comparatively massless against the energy of the process. 

Let us now characterize the backgrounds. The final state we are looking for is the same as a single top production with the top decaying leptonically except for the sign of the charge of the lepton relative to the one of the $b$-jet, that is not easy to measure. However, although irreducible, it is not a main background as it will be explained below. $W$ production associated with light jets, when the $W$ decays leptonically and one of the high $p_T$ light jets is mistagged as a $b$-jet, is background for this signal also. Despite the fact that the $b$-tag has a tiny contamination coming from light jets the cross-section for $W$ + light jets production is so huge that it becomes the largest background the signal has. Observe that the event topology of this background is much the same as the one of the signal. If we are looking for a high $p_T$ jet that is mistagged as a $b$-jet the $W$ will have high $p_T$ in the opposite direction because of the transverse momentum balance. This leads background events with the same kinematical features that the signal as we discussed above.  

In second place, the other main background is \ttbar , when only one of the top quarks decays leptonically and one bottom is missed. Although top quark pair production has larger cross-section than single top production, this is fairly not enough to explain why it is more important. This is so because a high $p_T(b)$ is more suppressed in the case of single top production than in pair production. Single top production is mainly through a $t$-channel exchanging a $W$ while \ttbar \ production is through $t$-channel exchanging a top and a $s$-channel. A particle coming from a $t$-channel production is more likely to be forward, and the tendency to this is increased as the mass of the exchanging particle is lower. Therefore, the $b$ which comes from the decay of the top in the single top production is expected to be more forward than in the case of pair production, resulting in events with lower $p_T(b)$.  
 
QCD multijet backgrounds, even with a mistagged $b$-jet and one fake lepton, could be important. The missing energy could come from an energy imbalance from the poor determination of the jet momentum. The larger is the jet transversal momentum the larger could be the fake missing energy. At high $p_T$, the measured jet momentum is within a 5 \% of the actual momentum of the jet at 1$\sigma$ \cite{ATLAS:2013wma}. Therefore, to avoid this background we require the minimum missing energy to be at least a 20 \% of the minimum momentum of the $b$-jet, that will be the highest $p_T$ jet. Assuming a normal distribution for the measured jet momentum, this gives a loose 4$\sigma$ that leads a QCD multijet contribution of less than 1 event at 100 fb$^{-1}$ for $p_T(b) > 600$ GeV taking into account also the probability of a fake lepton and the mistagging rate of the $b$-jet algorithm. This estimate is enough for our purposes.   

Other backgrounds to this signal are $W$ + $b$ and $W$ + $b \bar b$ when one of the bottom is missed or $W$ + $c$ and $W$ + $c \bar c$ when one $c$ is mistagged. Also, $Z$ + light jets, $Z$ + $b$ and $Z$ + $b \bar b$ when $Z$ decays to charged leptons, and one of them is missed. We have found all these backgrounds to be negligible, so that in the following we only show $W$ + light jets and \ttbar \ backgrounds.

Throughout this work we have considered a large region of the parameter space of masses and couplings, with $g_R\leq0.065$ and $1.3\ {\rm TeV}\leq m_V\leq2.5\ {\rm TeV}$, reporting our results for that set in Fig.~\ref{reach}. We have chosen two signal benchmark points for the plots, tables and optimization of the cuts: a reference point 1 of coupling $g_R = 0.035$ and mass of 1.3 TeV and a reference point 2 with $g_R = 0.046$ and mass of 1.8 TeV.  In the first row of Table \ref{table1} we show the total NLO cross-section for the reference points and the main backgrounds. Although in the single production dominated region we have a statistically significant number of events for the signal at 100 fb$^{-1}$, the background are huge and we have to design cuts to show up the signal over the background fluctuations.

\begin{table}
\begin{center}
\begin{tabular}{|c|c|c|c|c|c|c|c|c|c|}
\hline
\multirow{2}{*}{$N_{\text{lep}}$} & \multirow{2}{*}{$N_{b\text{-jet}}$} & \multirow{2}{*}{$N_{\text{jet}}$} & $\sigma(t\bar t)$ & $\sigma$($W$ + jets) & $\sigma(S_1)$ & $\sigma(S_2)$  \\
& & & (pb) & (pb) & (pb) & (pb) \\
\hline
1 & - & - & 148 & 7400 & 3.4 $\times$ 10$^{-3}$ & 1.8 $\times$ 10$^{-3}$ \\
1 & 1  & - & 57.3 & 42.1 & 1.3 $\times$ 10$^{-3}$ & 0.65 $\times$ 10$^{-3}$ \\ 	 	
1 & 1  & 1 or 2 & 15.4 & 18.9 & 0.84 $\times$ 10$^{-3}$ & 0.42 $\times$ 10$^{-3}$  \\	
\hline
\end{tabular}
\end{center}
\caption{\label{table1} Single production cross-section for the two signal reference points and main backgrounds in the one lepton channel, one $b$-tagged jet and after jet multiplicity cuts.}
\end{table}

We have simulated signal and background for LHC at 14 TeV with \texttt{MadGraph/MadEvent 5} \cite{mgme}. We pass them to \texttt{Pythia 6} \cite{pythia} for showering and hadronization and to \texttt{PGS} \cite{pgs} for detector simulation. The jets are reconstructed using the anti-$k_T$ algorithm with $R = 0.4$ provided by \texttt{PGS}. We have used NLO K-factor of Ref.~\cite{Campbell:2009gj} for the normalization of the signal and the ones of Ref.~\cite{Huston:2010xp,Cacciari:2008zb} for the backgrounds. Before the analysis we apply usual pre-selection cuts for the LHC. For charge leptons we require $p_T(\ell) > 25$ GeV and $\eta(\ell) < 2.5$. For reconstructed jets we require $p_T(j) > 25$ GeV and $\eta(j) < 4$.  Finally, we required for the missing energy $\met > 25$ GeV.  
We used PGS with the original tune, that for $b$-tagging in the high $p_T$ regime has an efficiency of about 40 \% and mistagging of 0.5 \%. Notice as reference that in Ref.~\cite{Chatrchyan:2012jua} the CMS experiment has reported for $p_T > 500$ GeV an efficiency of about 55 \% and mistagging rates of about 3 \% for light quarks jets, showing consistency of our working point. The $b$-tag algorithm works for jets within $\eta(j) < 2.5$ and all $b$-jets out of that region are considered as light jets.  
\begin{figure}[!htb]
\begin{minipage}[b]{0.5\textwidth}                                
\begin{center}          
\includegraphics[width=0.9\textwidth]{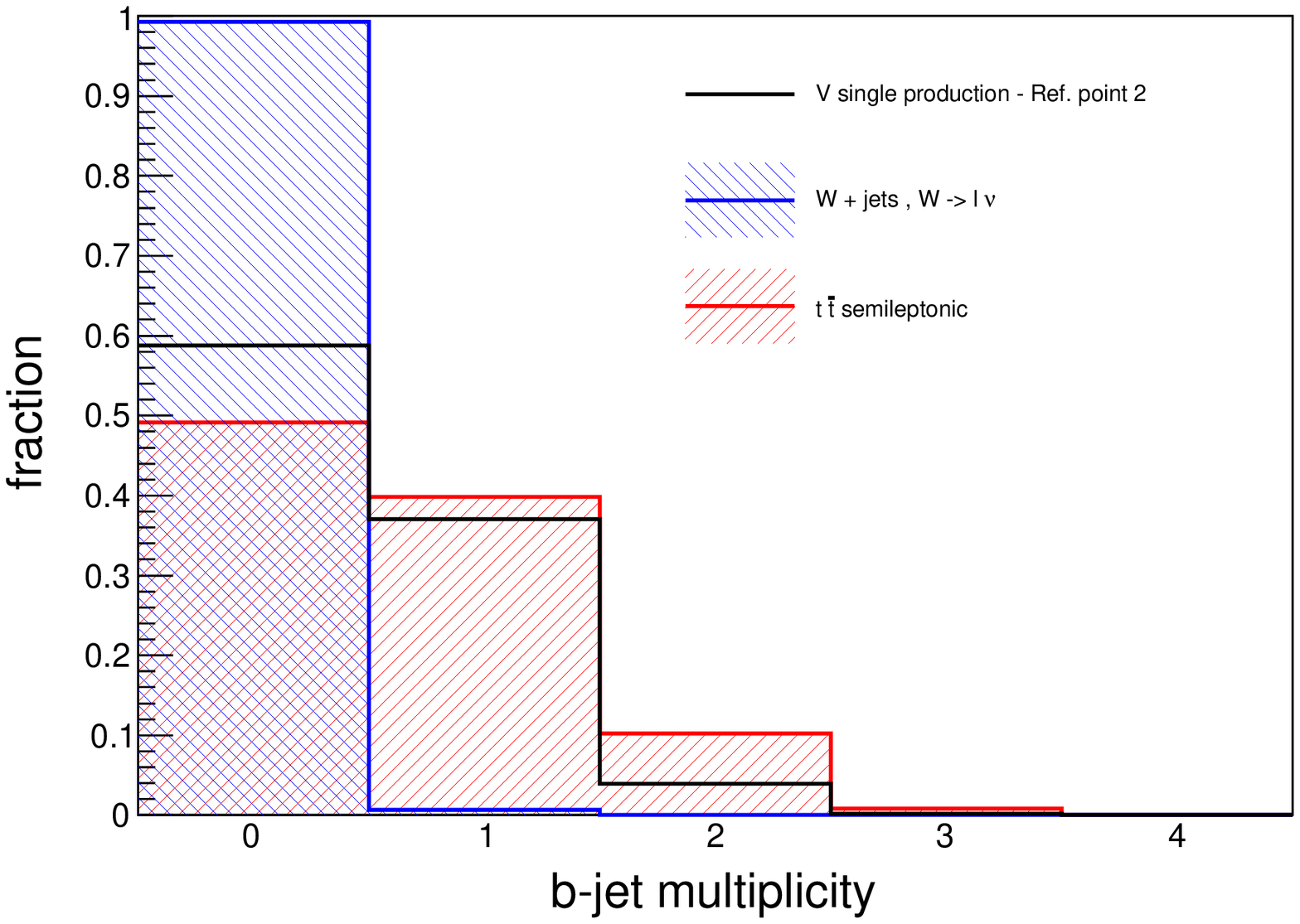}
\newline
(a)
\end{center}
\end{minipage}	
\begin{minipage}[b]{0.5\textwidth}                                
\begin{center}                                                           
\includegraphics[width=0.9\textwidth]{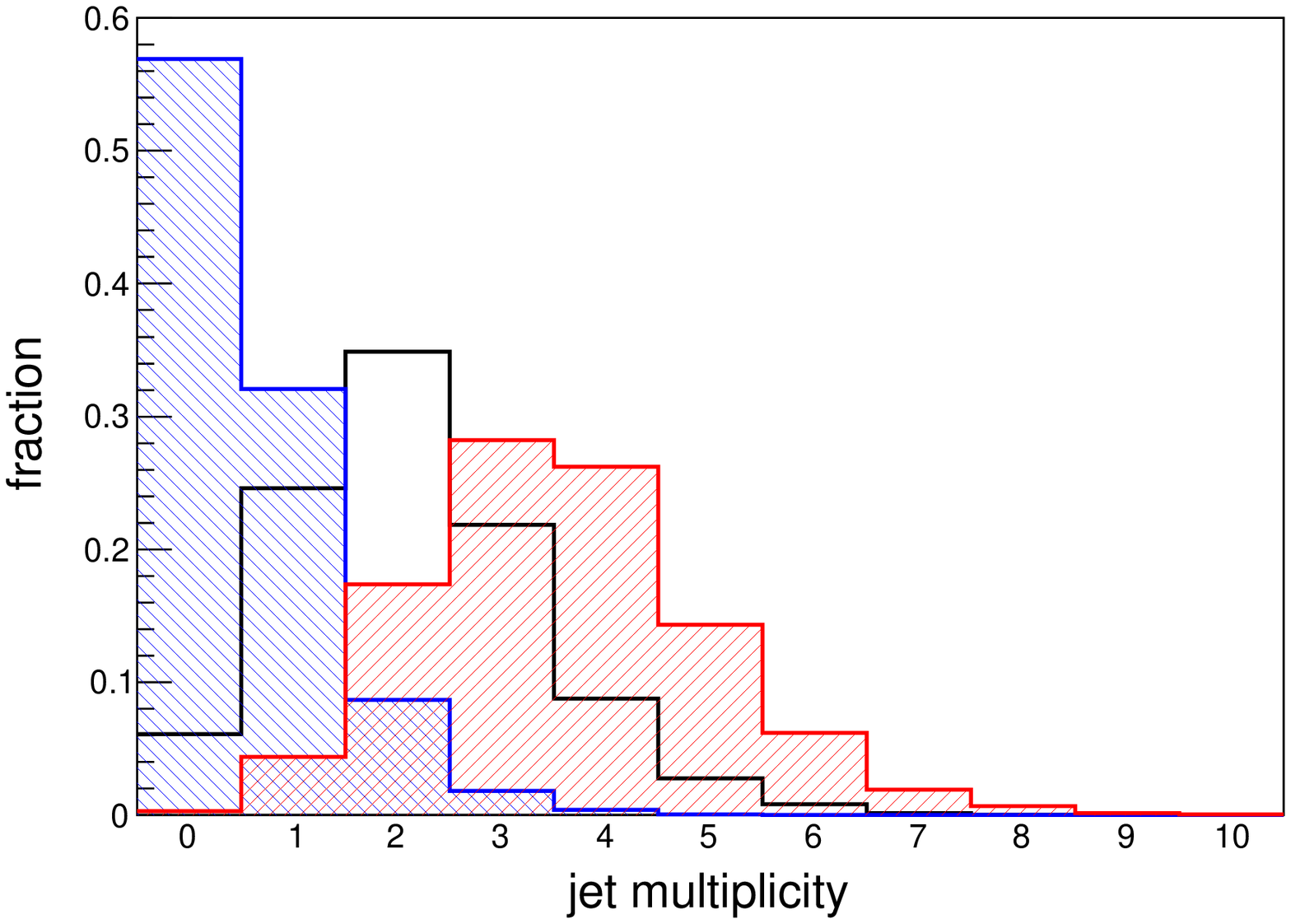}
\newline
(b)
\end{center}
\end{minipage}
\caption{\label{multiplicities}[color online] (a) $b$-jet and (b) jet multiplicity for the benchmark signal $m_V = 1.8$ TeV and $g_R = 0.046$ and main backgrounds. The signal for $m_V=1.3$ TeV is similar to 1.8 TeV. The jet multiplicity is shown having already asked for one $b$-jet.} 
\end{figure}

We study in first place variables that define the final state as $b$-jet and jet multiplicities. After that we discuss kinematical variables that can lead to an enhancement of the signal over the background. 

In Fig.~\ref{multiplicities} (a) we show the $b$-jet multiplicity of the benchmark signal and main backgrounds. As it was expected, asking for more than one $b$-jet rejects most of the $W$ + jet background. The initial $b$-quark for the signal necessarily comes from a gluon splitting. Therefore, along with the $b$ quark coming from the decay of the $V$ there is a $b$ quark from initial state radiation. Thus, although both, the signal process and \ttbar \ have two $b$-jets, notice that the rate of missed second $b$-jet is higher in the case of the signal because most of the initial state $b$-quarks are forward and escape from the $b$-tagger. However, in the practice there is a slight difference between asking for exactly one or more than one $b$-jet. We choose the former. In Table \ref{table1} we show the cross-section for benchmark signal and main backgrounds after select events with exactly one $b$-tagged jet.

The second feature we can notice is that the number of reconstructed jets has differences between the benchmark signal and background as we show in the Fig.~\ref{multiplicities} (b) after we have asked for exactly one $b$-jet. $W$ + jets background has the most contribution for low jet multiplicity, that is no jet or 1 jet.\footnote{Notice that we are calling jets only to light jets, so that taking into account the jet that was mistagged most of the background comes from $W$ + 2 to 3 jets.} Therefore, to suppress this background we require at least 1 jet. On the other hand, \ttbar \ background events has the highest number of jets, mostly between 2 and 4 jets, due to the hadronically decaying top. To suppress these backgrounds we ask for 2 jets at most. In Table \ref{table1} we show the cross-section for benchmark signal and main backgrounds after asking for 1 or 2 jets. 
\begin{figure}[!htb]
\begin{minipage}[b]{0.5\textwidth}                                
\begin{center}                                                       
\includegraphics[width=0.9\textwidth]{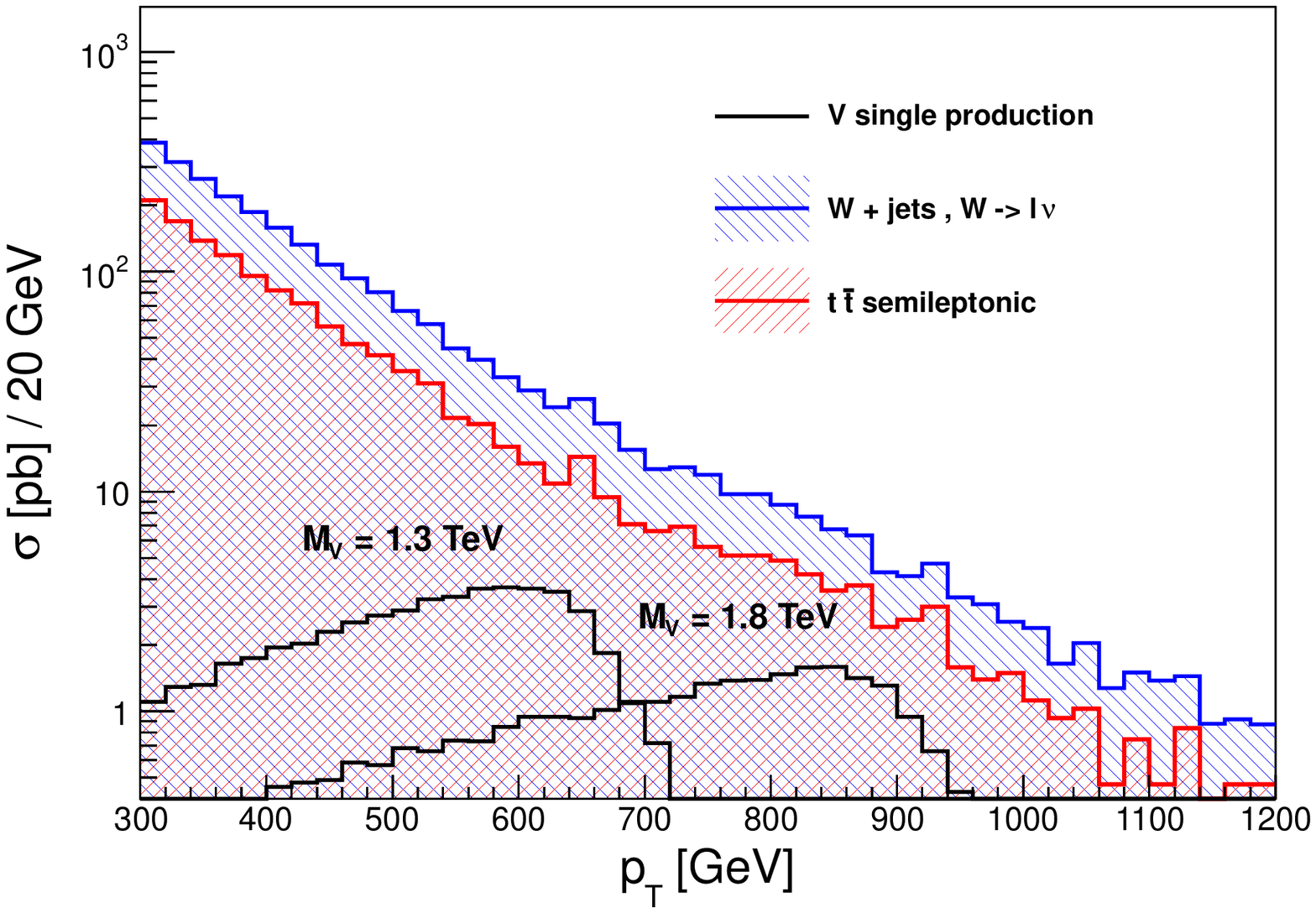}
\newline
(a)
\end{center}
\end{minipage}	
\begin{minipage}[b]{0.5\textwidth}                                
\begin{center}                                                                                                       
\includegraphics[width=0.9\textwidth]{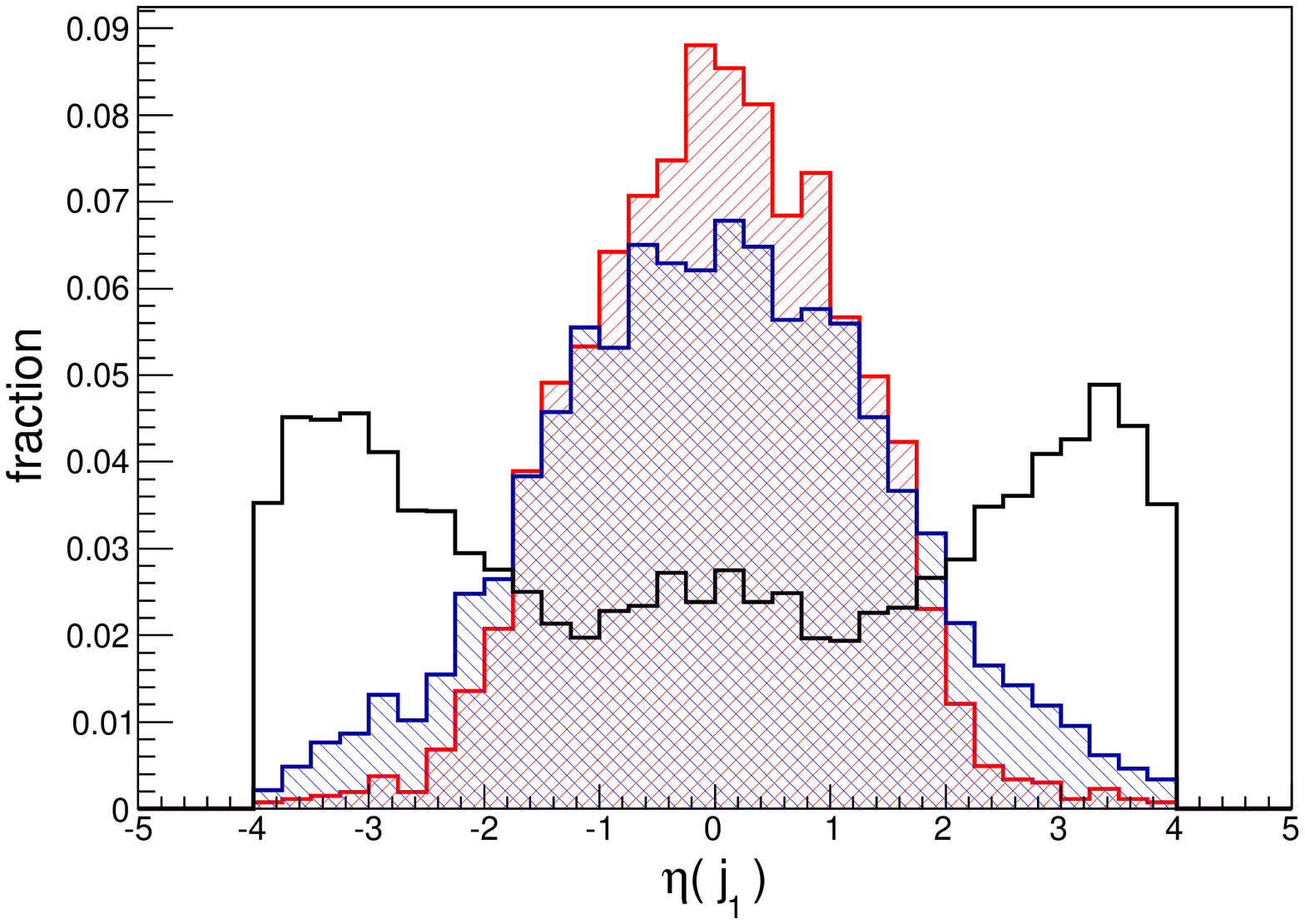}
\newline
(b)
\end{center}
\end{minipage}
\caption{\label{ptbandeta}[color online] (a) $p_T(b)$ differential cross-section for the two signal reference points and main backgrounds after $b$-jet and jet multiplicity cuts. (b) $\eta(j_1)$ distribution for signal $m_V = 1.8$ TeV and $g_R = 0.046$ and main backgrounds after the previous cuts and  $p_T(b) > 500$ GeV. The signal for $m_V=1.3$ TeV is similar to 1.8 TeV.} 
\end{figure}

As we discussed above, other useful variable is $p_T(b)$. This variable is highly efficient to enhance the signal, as we can see in Fig.~\ref{ptbandeta} (a). There, we have plotted the main backgrounds and the two benchmark signals after the previous cuts on $b$-jet and jet multiplicities. We can see that the $b$-jet takes half of the momentum of the V resonance and $p_T(b)$ is peaked around $m_V/2$.  

In Fig.~\ref{ptbandeta} (b) we have plotted the distribution of the rapidity of the leading jet ($\eta(j_1)$) for the benchmark signal of $m_V = 1.8$ TeV and the main background after the previous multiplicity cuts and after we have applied the cut $p_T(b) > 500$ GeV. Only after these cuts we can notice the accumulation of signal events in the forward region. The reason for this is as follows: in the case of the signal, the propagator of the $W$ boson in the $t$-channel becomes more peaked as the energy of the outgoing particles rise. On other hand, \ttbar \ background will be boosted after the cuts and one of the top quarks will be most likely to be in the central region because of the $\eta$ restriction on the $b$-tagger. Therefore, because of the momentum balance of the event the leading jet will come from the decay product of the other top and should be mostly central. The same is for $W$ + jet background. If we asked for a large momentum and central $b$, it is more likely to find a central leading jet because of the momentum balance.  

Other useful kinematical variable would be the invariant mass of the system ($\ell$, $\nu$, $b$), M($\ell$, $\nu$, $b$), for which one needs to reconstruct the four momentum of the neutrino. The invariant mass M($\ell$, $\nu$, $b$) will be peaked around the mass of the particle V for the signal. The knowledge of the mass of the $W$ and the transversal momentum of the neutrino allow us to determine its longitudinal momentum through a quadratic equation. In case the discriminant is positive there are two real solutions, we will take the one with the smaller absolute value for the longitudinal momentum. If the discriminant is negative there are two imaginary solutions. In this case we do not use any of the two solutions but we take the $\eta$ of the neutrino as the $\eta$ of the lepton. This solution becomes a better approximation for boosted and high invariant mass events. The $W$ + jets background can give large M($\ell$, $\nu$, $b$) when the $W$ and the jet which is mistagged are produced back-to-back and with high $p_T$. In the case of \ttbar \ production, the main contribution for high invariant masses is when the $b$-tagged jet is the one from the hadronically decaying top quark, because if it is from the leptonically decaying top quark it would reconstruct the mass of the top quark. We do not use the invariant mass as a cut variable because of the possible large systematic uncertainties in the determination of the neutrino momentum. The introduction of these systematic uncertainties requires a careful analysis that is beyond the scope of this work. However, we will show the invariant mass M($\ell$, $\nu$, $b$) distribution at various stages along this work to get an idea of the cuts' effects on it.   

\subsection{Cut scanning}

We have found that the best final state to find the signal is to ask for 1 lepton, 1 $b$-jet, missing energy and 1 or 2 light jets, being the leading one forward. Now we use the remaining relevant cuts to optimize the search strategy in this final state and to show up the signal over the background. The optimized cuts will be a set of kinematical cuts that will depend on the mass of the new resonance. Despite the fact that the significance is reduced with smaller couplings, the optimized set of cuts will not depend on it because the signal cross-section scales with the coupling square for $g_R\gg g_L$. 

We have generated the signal and main backgrounds as we detailed in the previous section, but to improve the generation time for the backgrounds we implemented some cuts at the parton level. For the $W$ + jets background we asked for at least one high $p_T$ jet of 200 GeV (the one that would be mistagged as $b$-jet). For \ttbar, we cannot know \emph{a priori} which of the two $b$-jets will be the missed one, then we cannot ask for at least one high $p_T$ $b$-jet without losing events. Therefore, we have generated \ttbar \ with only one leptonically decaying top with $p_T(\ell)  > 100 $~GeV. We will show the results of the simulation for the early LHC run II of 100 fb$^{-1}$ although we used a generated sample of 1000 fb$^{-1}$ for backgrounds to reduce statistical fluctuations.   

We have scanned randomly over $p_T(b)$, $\eta(j_1)$, $\met$ and $p_T(\ell)$ in order to find the best cuts for both reference points with $m_V=1.3$ and $1.8$ TeV. The scan was over the whole allowed range for these variables. The only restriction arises from the cuts implemented at the generation level and the requirement that the missing energy cut is at least 20\% of the cut on the $b$-jet. In the case of $p_T(b)$ the best cut will be far beyond the limit of 200 GeV of the generation. In the case of $p_T(\ell)$ we cannot go below 100 GeV for the reasons explained in the previous paragraph, but we do not expect much improvement in that region. To calculate the significance we assume the signal and background events to follow a Poisson distribution. The $p$-value, {\it i.e.}: the probability to obtain at least as many signal events as $S$ with expected background $B$ is:    

\begin{equation}
p = \sum_{n=S+B}^\infty \frac{B^n e^{-B}}{n!}.  \label{pvalue}
\end{equation}
\begin{table}
\begin{center}
\begin{tabular}{|c|c|c||c|c|c||c|}
\hline
$p_T(b)$ & $\met$ & $\eta(j_1)$ & $\sigma(t \bar t)$ & $\sigma$($W$ + jets) & $\sigma(S_1)$ & significance  \\
 & & & (fb) & (fb) & (fb) & 100 fb$^{-1}$  \\
\hline
\hline
 300 GeV &  -      &  -   & 12.79 & 11.52 & 0.5 & 1 \\
 500 GeV &  -      &  -   & 2.47 & 2.4 & 0.31 & 1.4 \\
 500 GeV & 100 GeV &  -   & 2.17 & 1.84 & 0.29 & 1.42 \\
 500 GeV & 100 GeV & 2.5 & 0.05 & 0.19 & 0.15 & 2.72 \\ 
\hline
\end{tabular}
\end{center}
\caption{\label{table2} Optimized cuts for the reference point 1 with $m_V = 1.3$ TeV and $g_R = 0.035$ at 100 fb$^{-1}$. The $\met$ has been taken as at least 20\% of the $p_T(b)$ to suppress possible QCD background.  After all cuts the backgrounds events are 24 and the signal events 15. An overall cut $p_T(\ell)~>~100$~GeV has been applied.}
\end{table}
\begin{table}
\begin{center}
\begin{tabular}{|c|c|c||c|c|c||c|}
\hline
$p_T(b)$ & $\met$ & $\eta(j_1)$ & $\sigma(t \bar t)$ & $\sigma$($W$ + jets) & $\sigma(S_2)$ & significance  \\
 & & & (fb) & (fb) & (fb) & 100 fb$^{-1}$  \\
\hline
\hline
 300 GeV &  -      &  -   & 12.79 & 11.52 & 0.29 & 0.59 \\
 700 GeV &  -      &  -   & 0.67 & 0.62 & 0.17 & 1.41 \\
 700 GeV & 150 GeV &  -   & 0.55 & 0.42 & 0.15 & 1.46 \\
 700 GeV & 150 GeV & 2.5 & 0.013 & 0.047 & 0.078 & 2.63 \\ 
\hline
\end{tabular}
\end{center}
\caption{\label{table3} Optimized cuts for the reference point 1 with $m_V = 1.8$ TeV and $g_R = 0.046$ at 100 fb$^{-1}$. The $\met$ has been taken as at least 20\% of the $p_T(b)$ to suppress possible QCD background. After all cuts the backgrounds events are 6 and the signal events 8. An overall cut $p_T(\ell)~>~100$~GeV has been applied.}
\end{table}

\begin{figure}[!htb]
\begin{center}
\begin{minipage}[b]{0.45\textwidth}                                
\begin{center}                                                       
\includegraphics[width=0.9\textwidth]{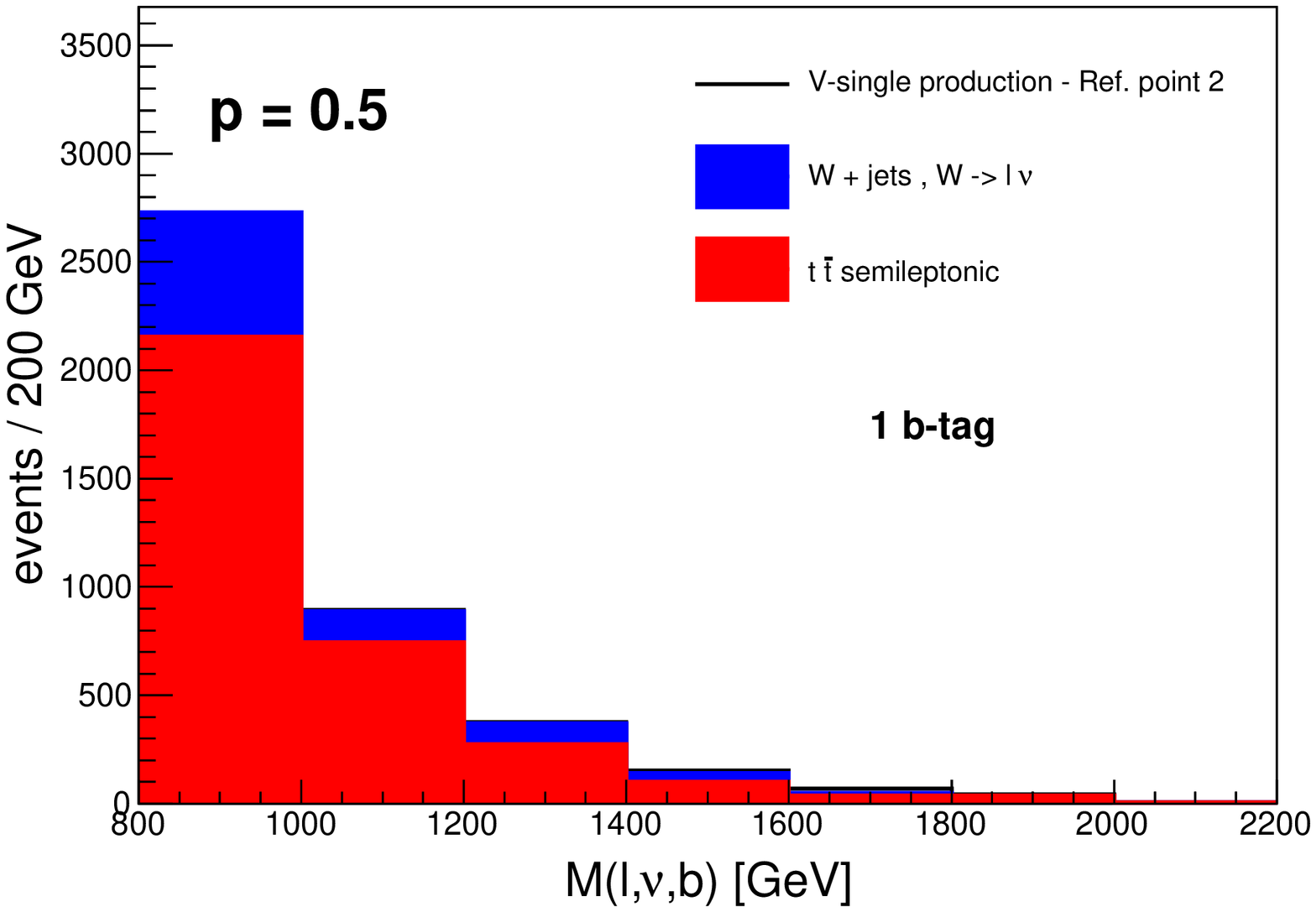}
\newline
(a)
\end{center}
\end{minipage}	
\begin{minipage}[b]{0.45\textwidth}                                
\begin{center}                                                                                                       
\includegraphics[width=0.9\textwidth]{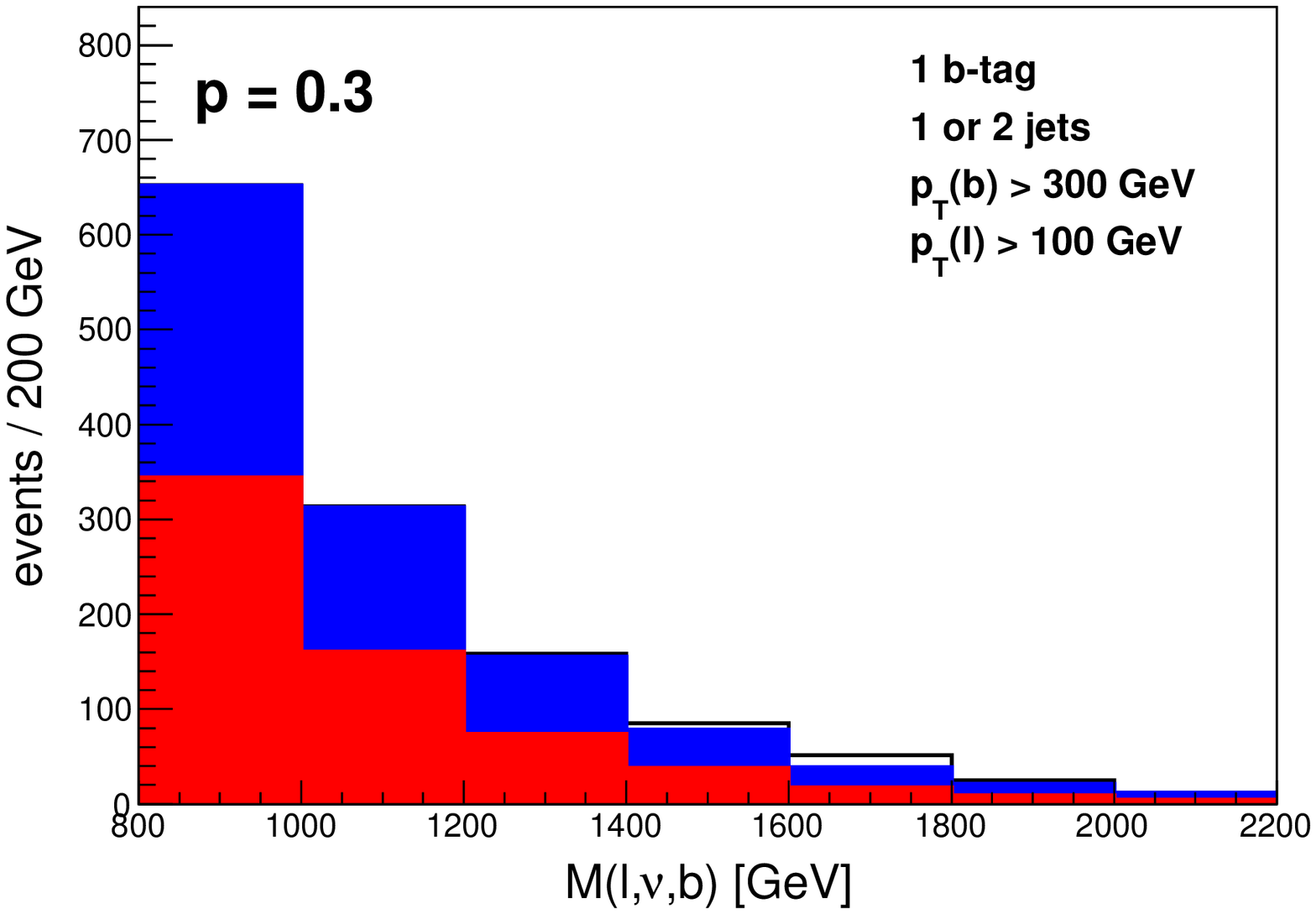}
\newline
(b)
\end{center}
\end{minipage}
\begin{minipage}[b]{0.45\textwidth}                                
\begin{center}                                                       
\includegraphics[width=0.9\textwidth]{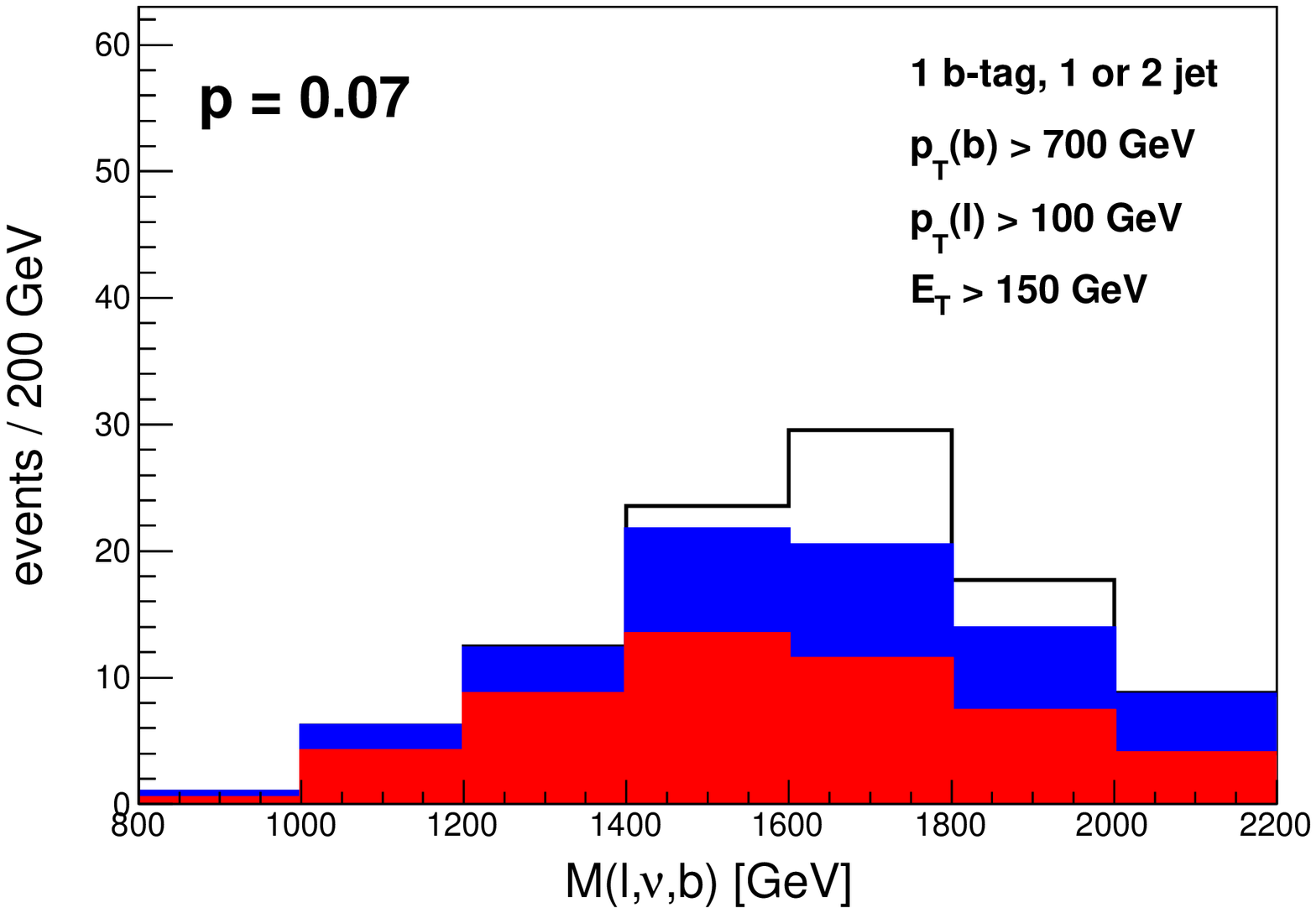}
\newline
(c)
\end{center}
\end{minipage}	
\begin{minipage}[b]{0.45\textwidth}                                
\begin{center}                                                                                                       
\includegraphics[width=0.9\textwidth]{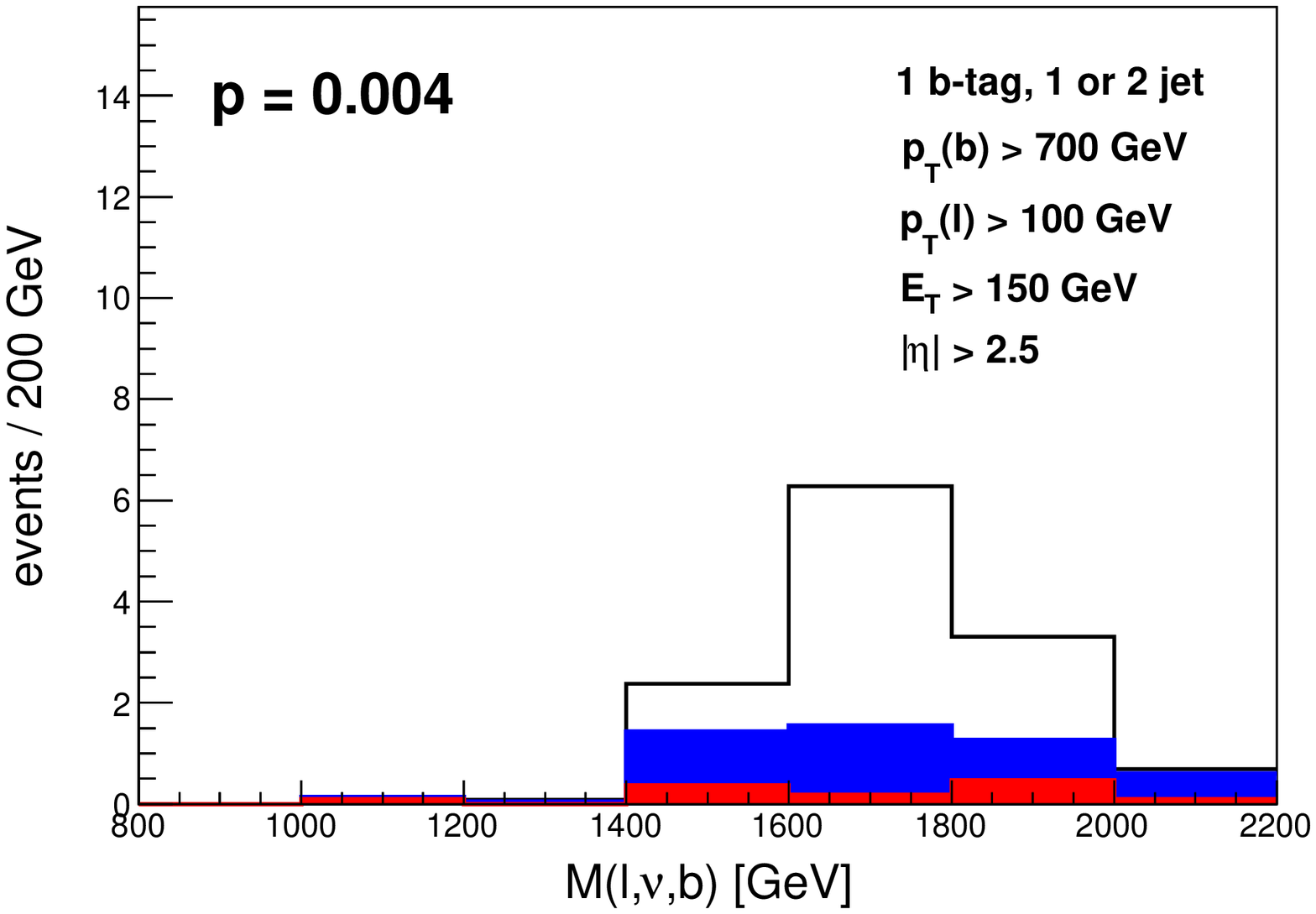}
\newline
(d)
\end{center}
\end{minipage}
\caption{\label{mtop}[color online] Invariant mass M($\ell$, $\nu$, $b$) after cuts for reference point 2 ($m_V = 1.8$ TeV and $g_R = 0.046$). } 
\end{center}
\end{figure}

In Tables \ref{table2} and \ref{table3} we show the optimized cuts for the reference points 1 and 2 respectively. To understand the role of the cuts we show how the signal and backgrounds are reduced as we apply each cut. An overall cut $p_T(\ell)~>~100$~GeV has been applied for the two reference points. We have not found an increment of the significance for more stringent cuts on $p_T(\ell)$. As expected, the $p_T(b)$ cut strongly reduces the backgrounds.  Notice in Table \ref{table2} (\ref{table3}) how, even after a strong pre-selection cut in $p_T(b)$, the optimal cut still reduces the backgrounds by a factor of 5 (20) while the signal is slightly reduced. The missing energy cut gives a tiny enhancement on the significance in both cases but a large QCD multijet background suppression which we are not showing, as discussed above. 
To illustrate further the effects of the cuts, we show in Fig.~\ref{mtop} the invariant mass M($\ell$,~$\nu$,~$b$) after the quoted cuts along with the $p$-value. 

The choice of the variables to include in the scan ($p_T(b)$, $\eta(j_1)$, $\met$ and $p_T(\ell)$) do not exhaust all the possibilities. One may choose to include M($\ell$, $\nu$, $b$) instead of $p_T(b)$ since they are correlated. In this case one can achieve a statistical significance slightly larger. However, as we discussed above, a realistic treatment of the invariant mass requires taking into account possible large systematic uncertainties. 

The key observable in this kind of single new resonance fermion production searches is the $p_T(b)$ which helps to isolate the signal for a wide range of the parameter space of the model as we will show in the next section. Because of this we have retained $p_T(b)$ as a cut variable instead of the more usual $H_T$\footnote{$H_T = \left|p_T(b)\right| + \left|p_T(\ell)\right| + \left|\met\right|$}. However, notice that there is no loss in generality because having fixed $p_T(\ell) > 100$ GeV and $\met > p_T(b)/5$, a $p_T(b)$ cut is equivalent to require $H_T~>~6/5 \ p_T(b)~+~100$~GeV.         

Additionally, we have checked that some other typical variables that we could have included in the scan do not improve the search. For instance, as a heavy particle is created one can expect the signal to emit radiation of smaller $p_T$ than the backgrounds which have more energy available. However, since we have already asked high $p_T(b)$, that it is not the case. Therefore, the vetoes in the leading and second jets are useless in this case.     

\subsection{Discovery reach}
In this section we will explore the discovery reach of our search strategy. The best search strategy should depend on the resonance mass $m_V$, however we find that the optimized cuts for the two reference points described in the previous section are enough to cover a wide range of masses and couplings for 100 fb$^{-1}$. We also show how the reach evolves with more luminosity.

To analyze the discovery reach of our search strategy we apply the optimized cuts we have found in the previous section for the two reference points to different samples of the signal varying the mass of the particle. We need to know how the significance of the signal over the background changes with the coupling $WVb$. Since the signal cross-section scales with the coupling, we re-scale the results for each mass to take into account different couplings. 

\begin{figure}[!htb]
\begin{center}
\begin{minipage}[b]{0.45\textwidth}                                
\begin{center}                                                       
\includegraphics[width=0.9\textwidth]{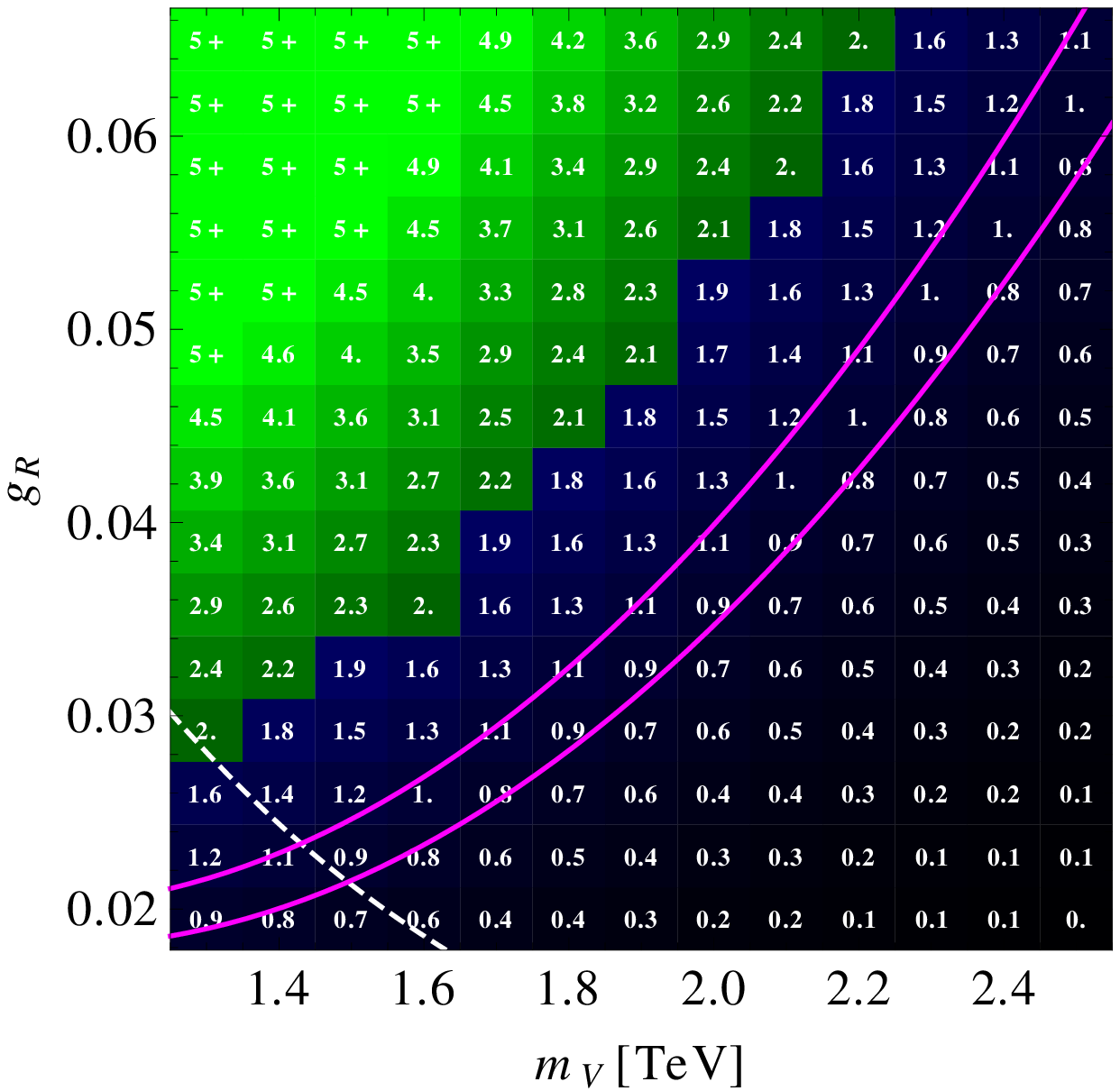}
\end{center}
\end{minipage}	
\begin{minipage}[b]{0.45\textwidth}                                
\begin{center}                                                                                                       
\includegraphics[width=0.9\textwidth]{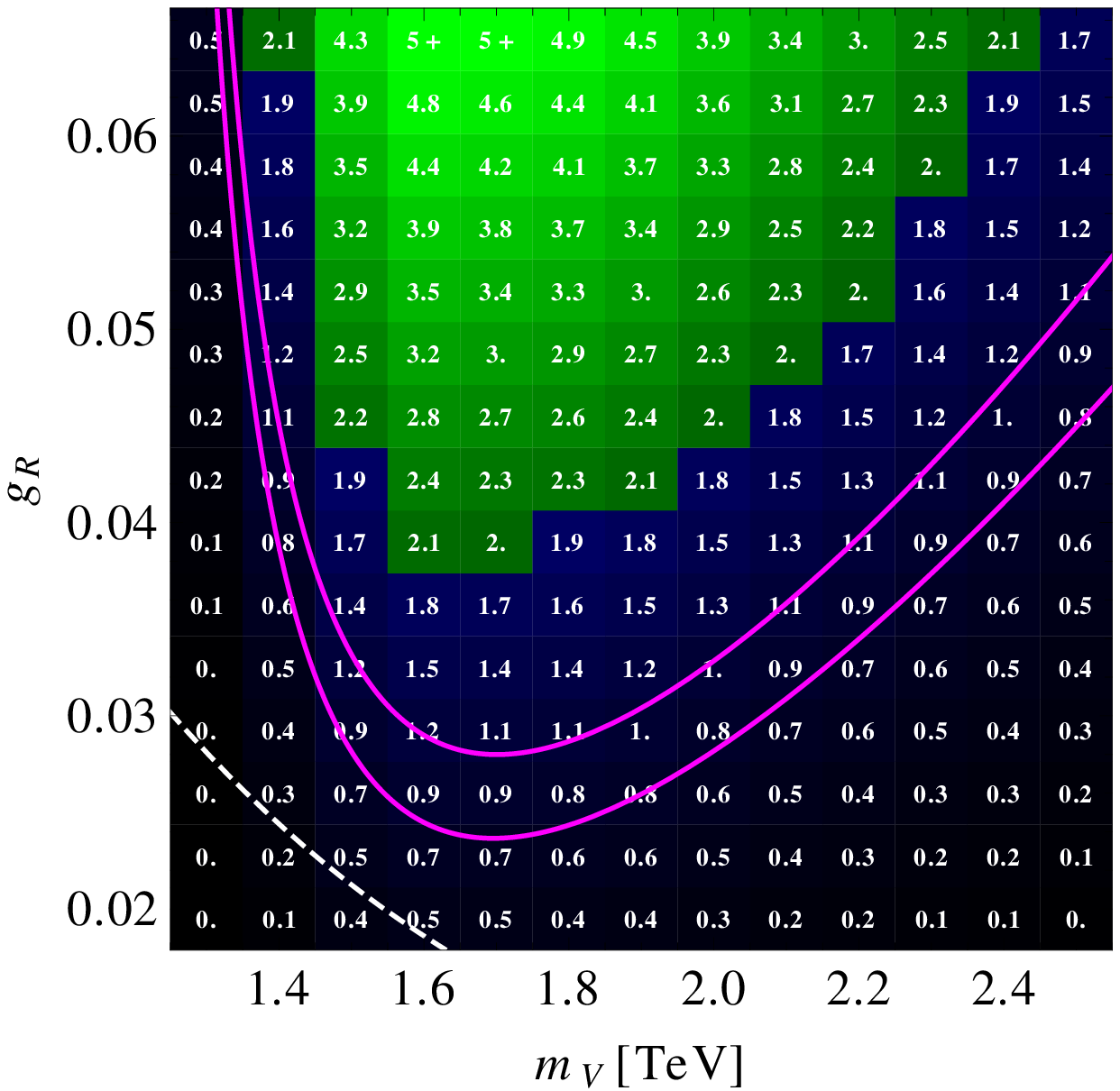}
\end{center} 
\end{minipage}
\vskip .6cm
\begin{minipage}[b]{0.45\textwidth}                                
\begin{center}                                                       
\includegraphics[width=0.9\textwidth]{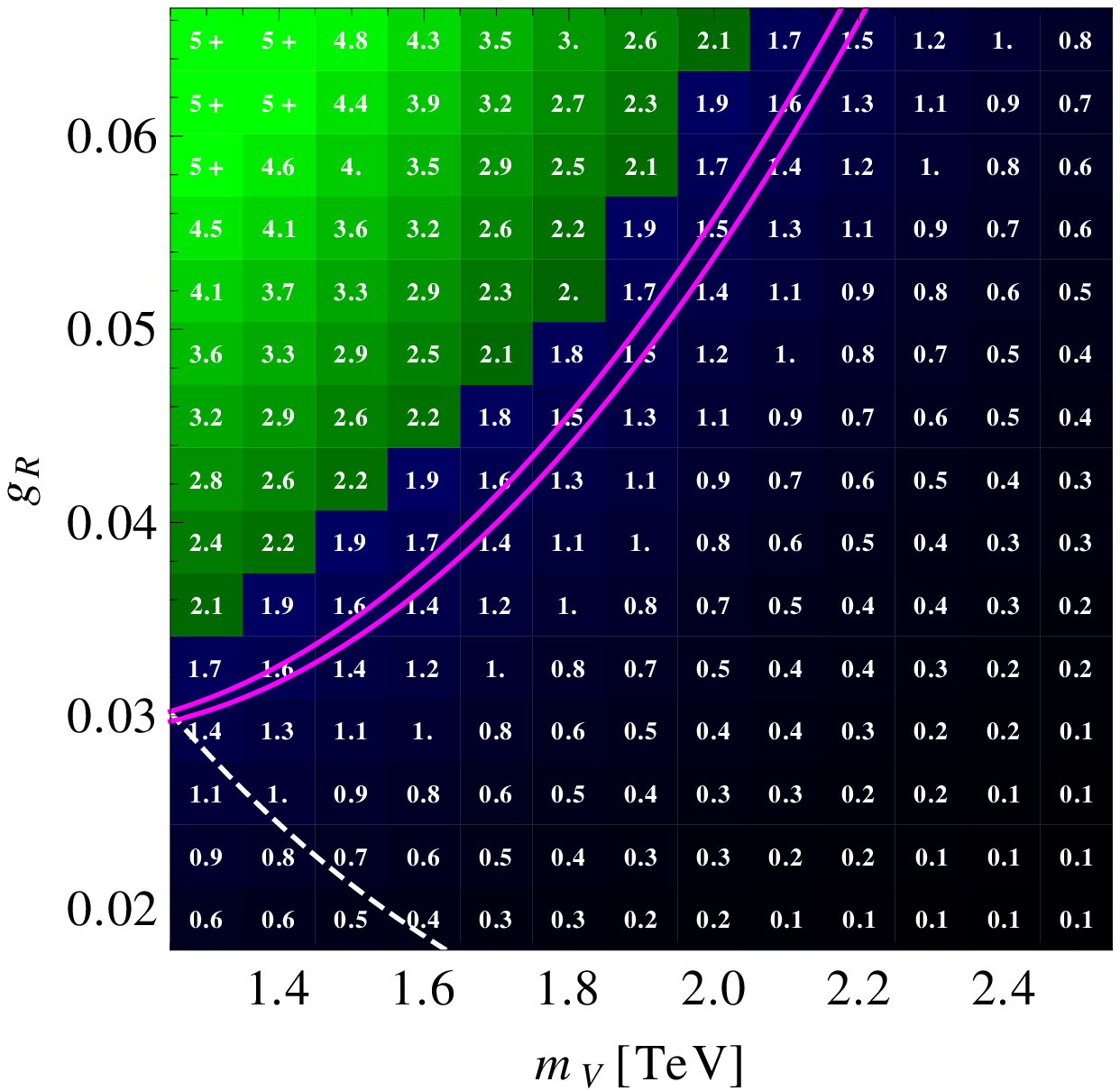}
\end{center}
\end{minipage}	
\begin{minipage}[b]{0.45\textwidth}                                
\begin{center}                                                                                                       
\includegraphics[width=0.9\textwidth]{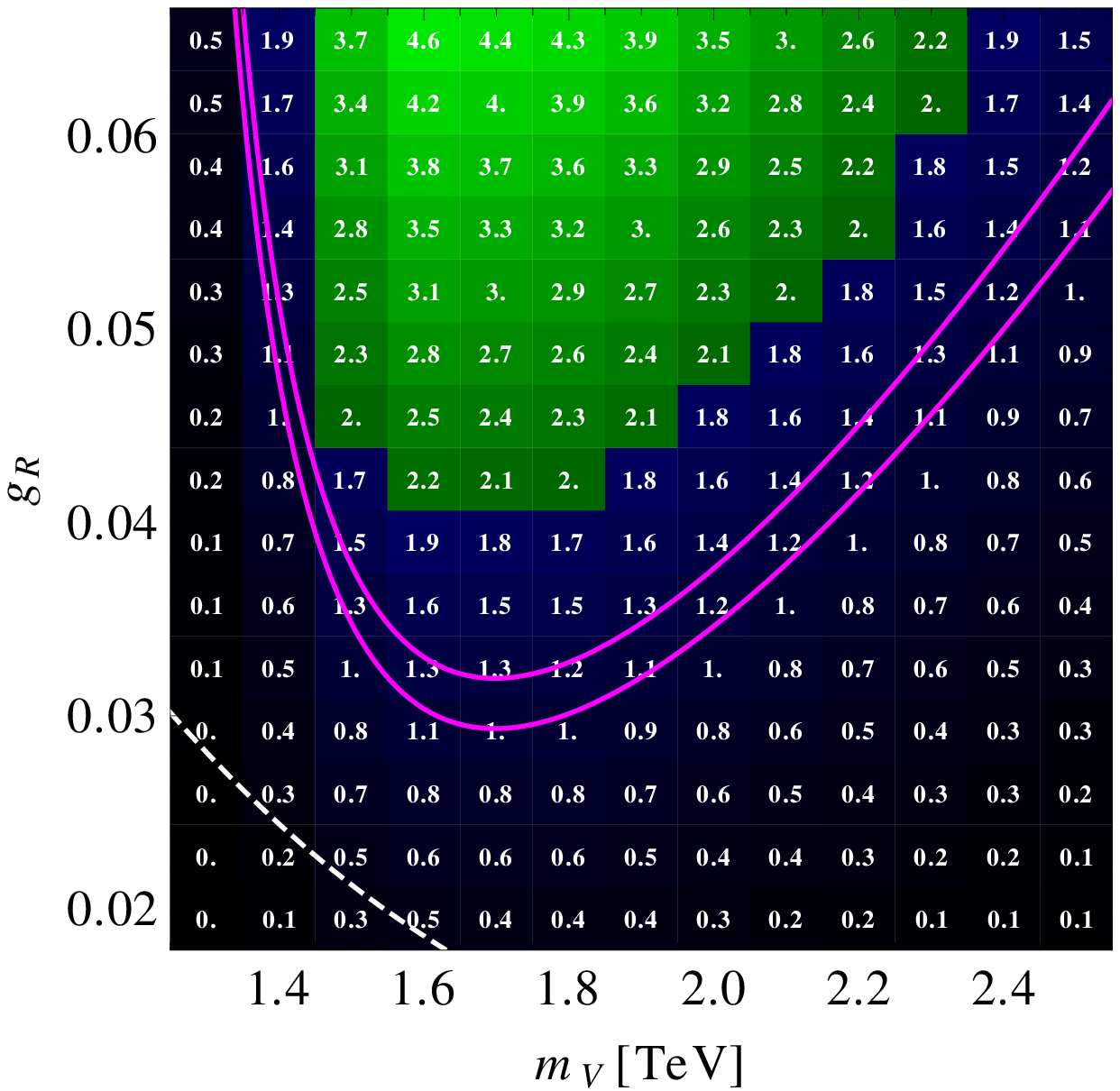}
\end{center} 
\end{minipage}

\caption{\label{reach}[color online] Top row: Statistical significance at 100 fb$^{-1}$ for different parameters $m_V$ and $g_R$ of the model for the optimized cuts for reference points 1 (left) and 2 (right). Bottom row: Full statistical and systematic significance at 100 fb$^{-1}$ for the same reference points, where systematic uncertainties are modeled as a 20 \% of the events. Green regions correspond to significance $>2$ and blue regions to significance $<2$. Below the dashed white line is the region where the cross-section for double production of V is larger than the single production. The magenta solid lines are the 2$\sigma$ contour lines for 300 and 500 fb$^{-1}$. } 
\end{center}
\end{figure}

In the top row of Fig.~\ref{reach} we show the significance in the plane $m_V$ vs.~$g_R$ for the two reference points at 100 fb$^{-1}$. For the best cut associated to the reference point 1 we have found that one can claim an evidence for a 2$\sigma$ discovery for masses up to 2.2 TeV for $g_R = 0.065$. Also, with that cut one can reach couplings as low as about $g_R = 0.035$ for masses up to a 1.6 TeV and $0.03$ for 1.3 TeV. For the best cut associated to the reference point 2 the reached mass is increased up to 2.4 TeV for $g_R=0.065$, while the coupling reach is $g_R = 0.04$ for 1.7 TeV. Notice that the background events only depend on the cuts and these are 24(6) for the best cut of point 1(2). Hence, the number of signal events for a 2$\sigma$ discovery is 11(6). This ensures that after applying the cuts the minimum number events is larger than 5 for all the region of masses and coupling of interest. 

The dashed white line in Fig.~\ref{reach} is where the cross-section for double and single production of $V$ are equal. Below the line usual pair production searches could be more useful although a precise determination requires the comparison of the efficiency of each search. In any case, the dashed white line provides an estimate of how this search strategy is complementary to the one for double production. We can see from Fig.~\ref{reach} that a considerable parameter space remains inaccessible between the reach for 100 fb$^{-1}$   and the white dashed line. This region can be probed with more luminosity. To see this, we also show in the Fig.~\ref{reach} two magenta solid lines corresponding to the contour lines of 2$\sigma$ for 300 and 500 fb$^{-1}$.

These results justify the choice of the reference point 1 and 2 as benchmark signal points. As we can see the optimized cut for 1.8 TeV are useful to claim a 2$\sigma$ evidence for all the range of masses between $1.4$ and $2.4$ TeV when $g_R=0.065$. But for the region below $1.5$-$1.6$ TeV and to probe all the range of coupling for $1.3$-$1.4$ TeV, it is necessary a second set of cuts, i.e. for 1.3 TeV.

\begin{figure}[!htb]
\begin{center}
\begin{minipage}[b]{0.45\textwidth}
\begin{center}
\includegraphics[width=0.9\textwidth]{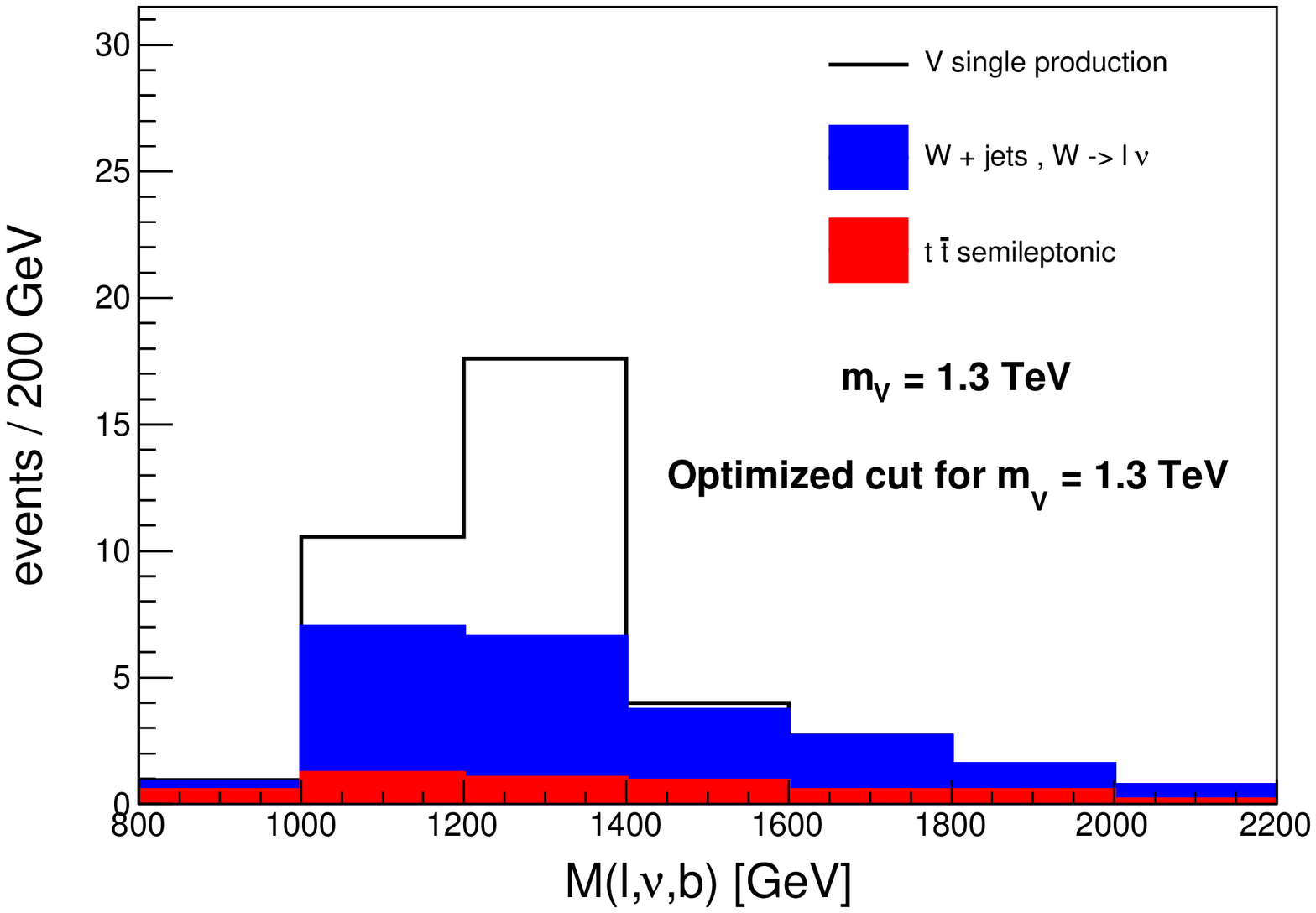}
\newline
(a)
\end{center}
\end{minipage}
\begin{minipage}[b]{0.45\textwidth}
\begin{center}
\includegraphics[width=0.9\textwidth]{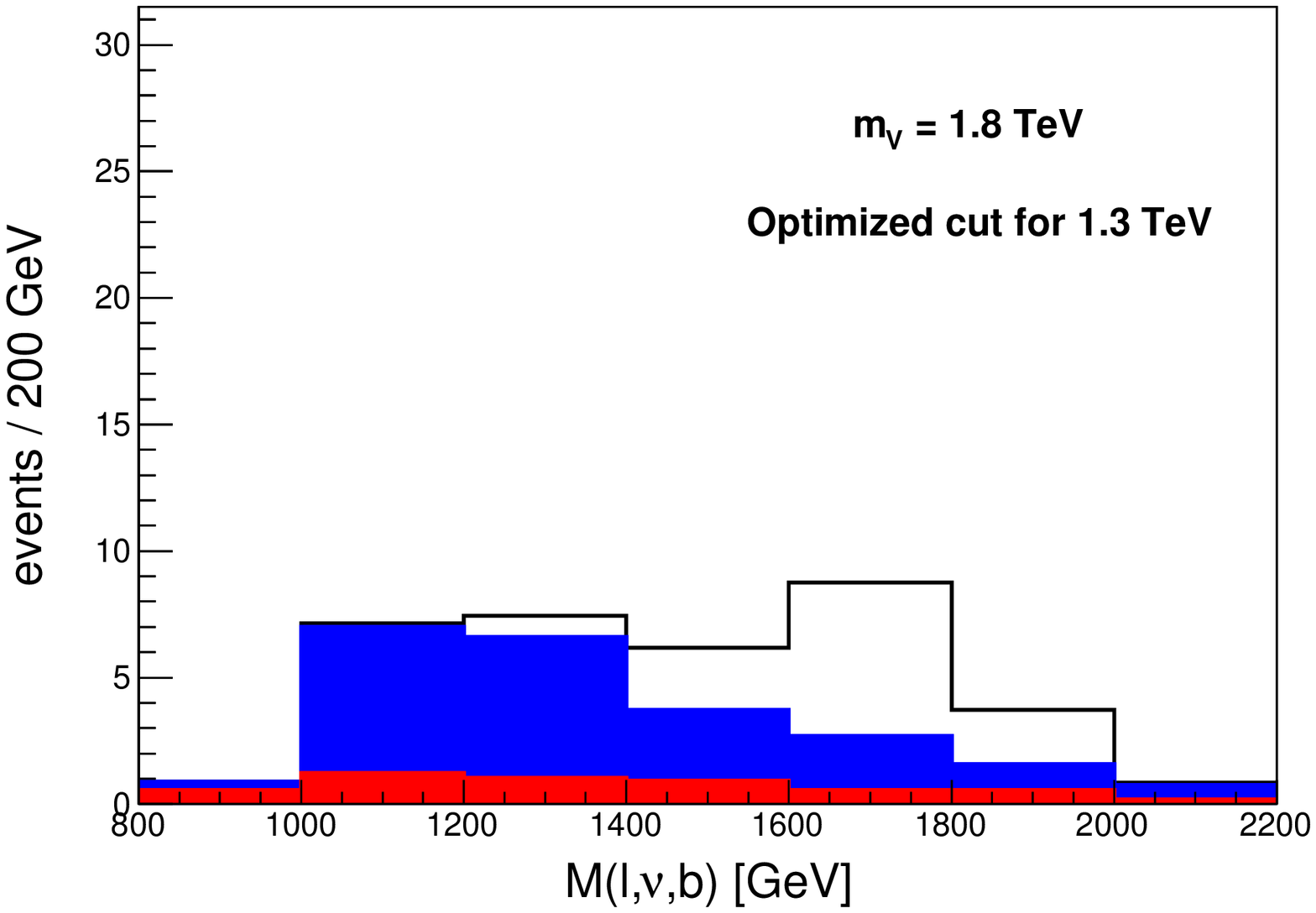}
\newline
(b)
\end{center}
\end{minipage}
\begin{minipage}[b]{0.45\textwidth}
\begin{center}
\includegraphics[width=0.9\textwidth]{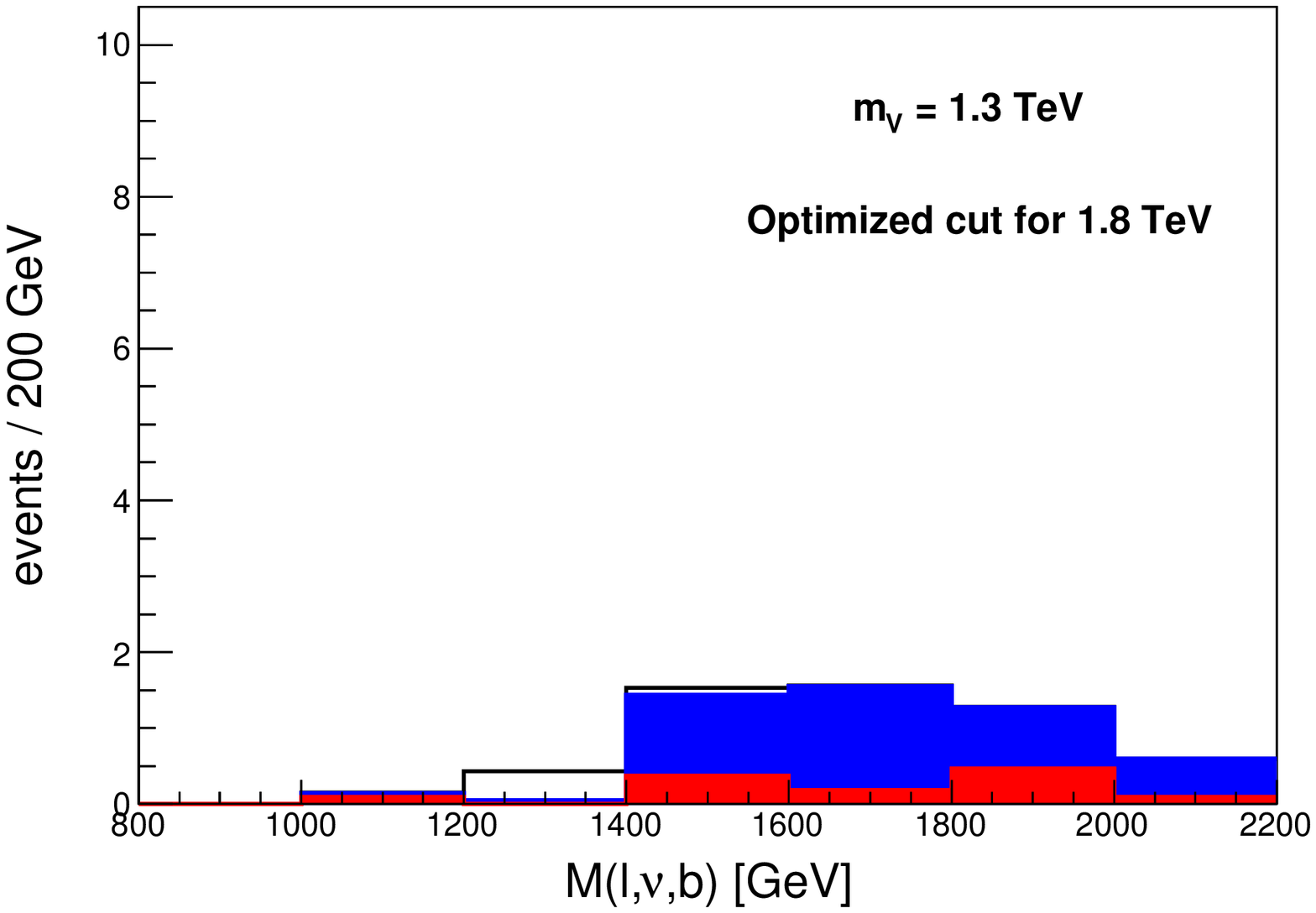}
\newline
(c)
\end{center}
\end{minipage}
\begin{minipage}[b]{0.45\textwidth}
\begin{center}
\includegraphics[width=0.9\textwidth]{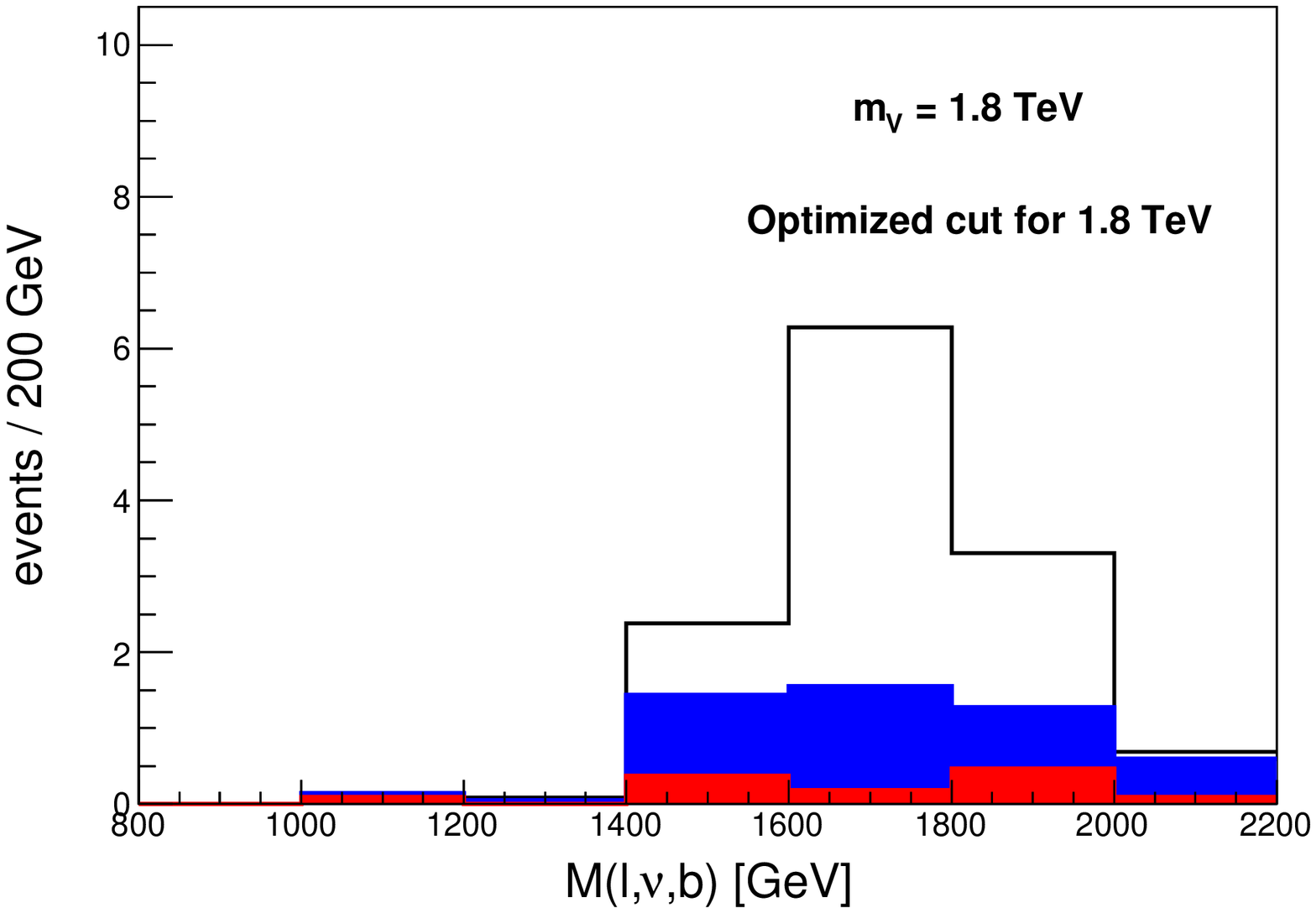}
\newline
(d)
\end{center}
\end{minipage}
\caption{\label{mtop2}[color online] Comparison of the invariant mass M($\ell$, $\nu$, $b$) for different signals and cuts to illustrate the possibility to obtain the mass of the particle after the strong cuts. In the first(second) column we show the invariant mass M($\ell$, $\nu$, $b$) for the reference point 1(2) after the cut optimized for that reference and for the other reference point. In plots (a) and (d) we see that, despite the strong cuts the peak in the signal corresponds to the mass of the particle.  In plots (b) and (c) we see that different designed cuts for different signals are required.}
\end{center}
\end{figure}

The second step, after finding an excess over the SM, would be to determine the properties of the new particle, as: charge, mass, spin. To estimate the quality of the invariant mass as an approximation for the mass of the new particle after the strong cuts we have applied, we plotted the invariant mass distribution in Fig.~\ref{mtop2}. This shows how two different signal behave after the same cut. In each row, we can see the effects of the cuts in the two reference points.  We can see that different optimized cuts for each signal are required to see an effect in the invariant mass distribution.  Moreover, it is worth noticing how after the strong cuts there is not an appreciable bias and the events of the signal are peaked around the mass of the resonance in panels (a) and (d).

\section{Discussion}
We have presented several results and distributions under the hypothesis of statistical uncertainties only. We want to consider now the impact of the systematic uncertainties. We start assuming that we know $B$ in Eq.~(\ref{pvalue}) within a 20 \% as usual in new fermion searches \cite{systematics}. We include this information as a Gaussian Bayesian prior of media $B$ and standard deviation $0.2 B$. In Fig.~\ref{reach} we show the full statistical and systematic significance (bottom row) in the plane $g_R$ vs.~$m_V$ along with the statistical only significance (top row). With this simplified model for the systematic uncertainties we have found that the overall reach is little reduced for both reference points.  The reason for this is that, because of the small signal cross-section, the only-statistical scanning chooses final selection cuts with few background events (see tables \ref{table2} and \ref{table3}).  Therefore, the systematic uncertainty of 20 \% does not produce an important modification to the total uncertainty.   Moreover, in the case of reference point 1 the systematic and statistical uncertainties are approximately equal and, therefore, more luminosity produces little enlargement in the reach, as it can be seen in the bottom left panel of Fig.~\ref{reach}.  For this case, a new cut scanning including systematic uncertainties would improve the reach.

The quantitative results of this work depends on the $b$-tag algorithm for high $p_T$. More precisely, the results of the scan show that any added variable that suppresses even more the \ttbar \ background is in general useless. The reason is that as the main background is $W$ + jets the scan chooses to reduce this background even when one can achieve a stronger suppression in \ttbar. In the case that the $W$ + jets can be further suppressed, either because of a better rejection of light jets by the $b$-tag algorithm or for other reason, \ttbar \ will dominate requiring modifications of the search strategy. In the next paragraph we discuss how to reduce \ttbar \ background.

\begin{figure}[!htb]
\begin{center}
\begin{minipage}[b]{0.45\textwidth}                                
\begin{center}                                                       
\includegraphics[width=0.9\textwidth]{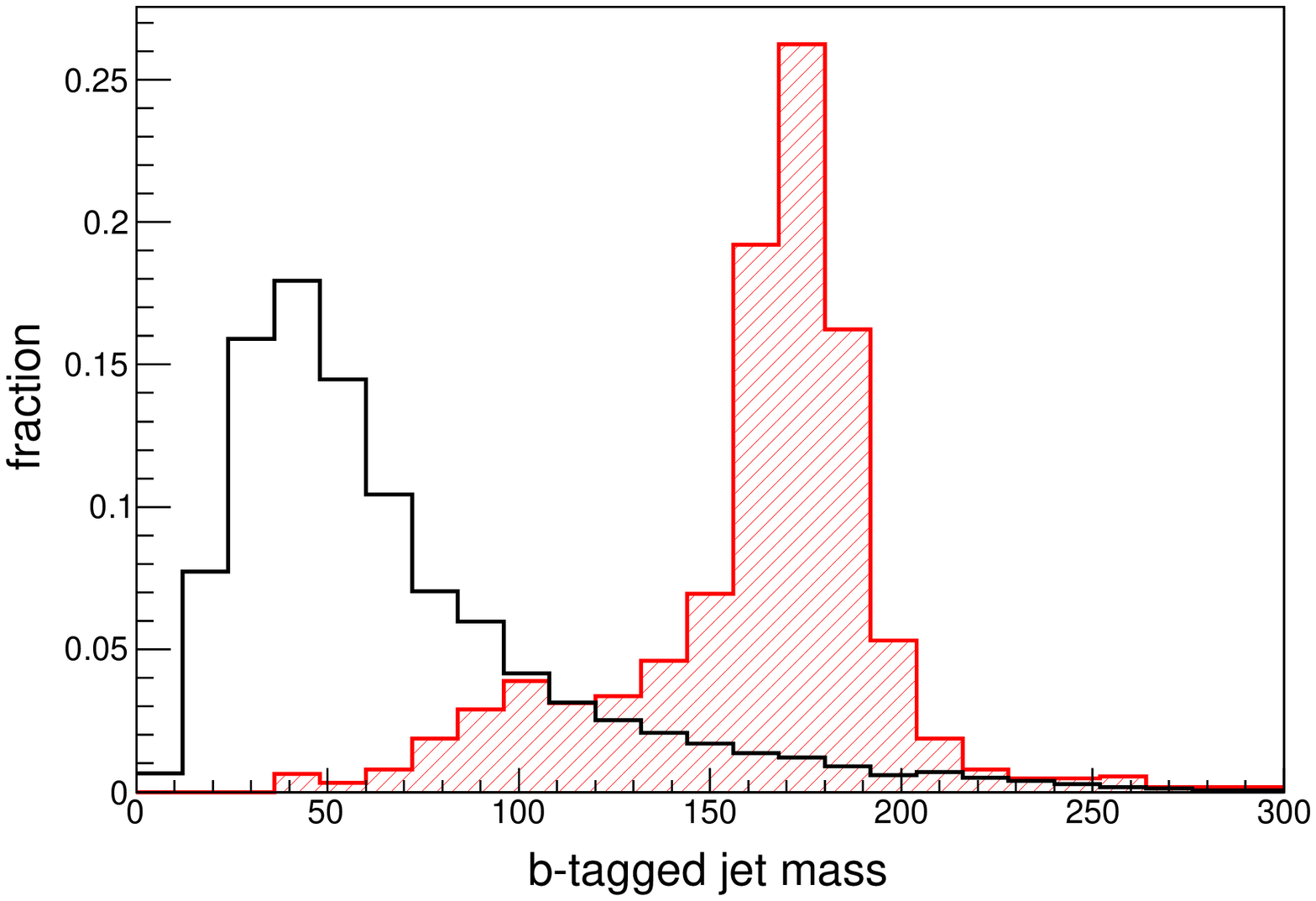}
\end{center}
\end{minipage}	
\begin{minipage}[b]{0.45\textwidth}                                
\begin{center}                                                                                                       
\includegraphics[width=0.9\textwidth]{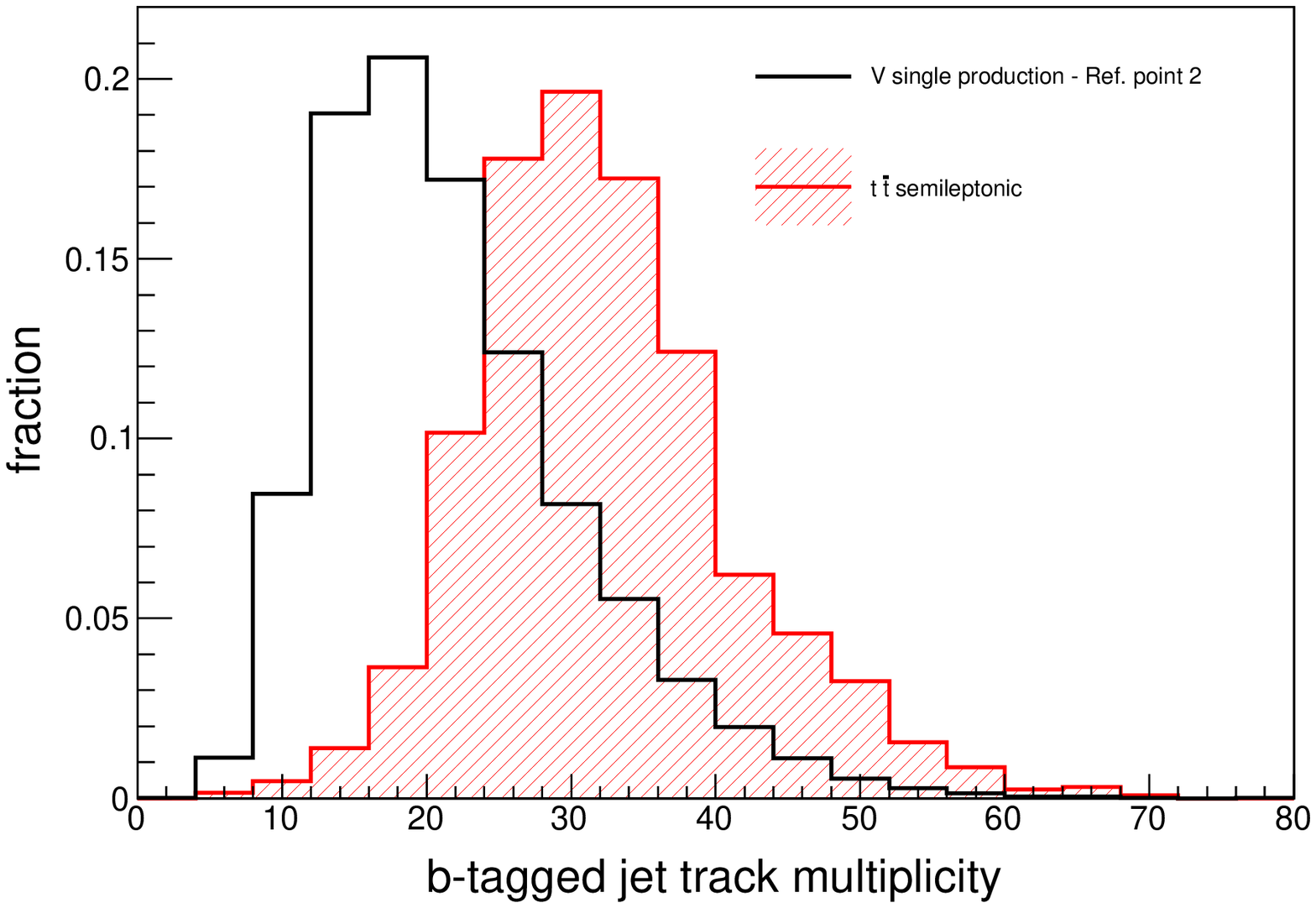}
\end{center}
\end{minipage}
\caption{\label{substructure}[color online] (Left) Mass of the reconstructed jet tagged as a $b$-jet for the reference point 2 and \ttbar \ background. (Right) Number of tracks in the reconstructed jet tagged as a $b$-jet for the reference point 2 and \ttbar \ background.  We have taken boosted events with M($\ell$, $\nu$, $b$) $> 1.5$ TeV and used anti-$k_T$ algorithm with $R = 0.6$ in order to reconstruct a jet which includes all the decay products of the hadronically decaying top quark.} 
\end{center}
\end{figure}

We have already observed that boosted \ttbar \ background events can be obtained only when the $b$-tagged jet is the one from the hadronically decaying top quark. But this means that the $b$-jet is likely to be part of a fat-jet that includes all the decays from the top quark. Then, a top tagger can be used rejecting events when one jet is tagged as a top. Without going into details we present two simpler jet substructure variables to illustrate the discriminating power between the signal and the \ttbar \ background. These variables attempt to expose the differences in the jet structure of the signal and \ttbar \ background. We have found that all the decay products of the hadronically decaying top quark are inside one fat-jet if we reconstruct jets with anti-$k_T$ and $R = 0.6$ for boosted events with M($\ell$, $\nu$, $b$) $> 1.5$ TeV. This is shown in Fig.~\ref{substructure} (a) where the reconstructed $b$-tagged jet includes most of the decay products of the top quark and reconstructs its mass. On the other hand, in Fig.~\ref{substructure} (b) we can see that the number of tracks is higher in the case of the \ttbar \ background because of the high activity in the fat-jet.

After an eventual discovery of a new fermion resonance, as any other particle, its properties should be determined. We have already shown how the mass of the particle could be measured, even after the strong cuts needed to look for a first evidence. The determination of the charge of the new particle will be more involved and will depend on the ability of the LHC to measure the charge of a $b$-jet. Measuring the sign of the $b$-jet charge relative to the $W$ charge it can be established if the new fermion has a charge 2/3 or -4/3. Notice that the ratios $\sigma_{\bar V}/\sigma_V \sim \sigma_{W^+}/\sigma_{W^-} \sim 2$ because of the proton composition. Hence, since $W+$jets is the main background, the lepton charge is not a good variable to discriminate signal and background.

Finally, the strength of the coupling $VWb$ could be determined with a measurement of the cross-section as usual. The V-A structure of this coupling could be tested with the $W$ polarisation. The $W$ polarisation is extracted from the angular distribution between the lepton and the $b$-jet in the $W$ rest frame. In the top quark SM, because of the Left nature of the EW coupling the $W$ Right-handed polarization is absent (only arise at NLO). Conversely, in our model the coupling is mainly right and the $W$ Left-handed polarization will be absent. As it can be easily deduced from angular momentum conservation, our right-handed signal angular distribution of the lepton and the $b$-jet will be sharply peaked at zero degrees.

\section{Conclusions}
We have considered a search for a bottom exotic partner that is complementary to the ones of top partners at LHC. Composite Higgs models aiming to solve the $A^b_{FB}$ anomaly measured at LEP and SLC generically require the presence of light partners of the bottom quark, including exotics fermions $V$ of charge -4/3. We have shown an effective theory with a composite Higgs and resonances where the correction to $A^b_{FB}$ is associated with the prediction of a light $V$-resonance. We also showed that partial compositeness predicts $m_V$ lighter than the composite scale, similar to the top partners. For masses larger than $\sim1$ TeV and couplings of order $g/10$ the model-independent pair production of the new particle is suppressed against the EW single production, favoring this signal at LHC. We have determined the typical size of couplings and masses within naturalness and have designed a search strategy for discovery over background
. We have made an approximate analysis of the discovery reach of this kind of particles at LHC with 14 TeV.

We have found that the best channels to find a signal of single production of this particle is to ask for one $b$-tagged jet, one lepton, missing energy and 1 to 2 jets. For this signal $W$ + jets and \ttbar \ production are the main backgrounds. We have designed a search strategy to show up the signal over the background and presented the optimized cuts for two reference points of the proposed model. The two sets of cuts, although optimized for two points only, are enough to cover a wide range of masses and couplings for 100 fb$^{-1}$ and also for higher luminosities. With them we have determined the range of masses and couplings that can be discovered at the LHC at 14 TeV. 

We have identified $p_T(b)$ and $\eta(j_1)$ as the key kinematical variables which help to enhance the signal over the \ttbar \ and $W$ + jet background. We used $p_T(\ell)$ and $\met$ to suppress the QCD background. We have performed a cut-based search strategy on these variables and found the optimal cuts to enhance the signal over the background. We have found that in the early LHC run II with 100 fb$^{-1}$ of integrated luminosity, the presence of this new particle can be tested up to masses of $2.4$ TeV with couplings of order $g/10$. Taking into account a systematic uncertainty of 20 \% the reach only drops to $2.3$ TeV. This reach covers a large region of the parameter space of a natural theory, aiming to solve the little hierarchy problem.

We have also found that substructure variables as the track multiplicity and the mass of the $b$-tagged jet can be used to discriminate between the signal and \ttbar \ background. \ttbar \ events produce a larger number of tracks than the signal, as well as larger masses for the reconstructed jets tagged as $b$-jet. This could be useful in case of improvement of the $b$-tagging algorithm in the high $p_T$ regime, so that the main background were \ttbar \ production. 

Finally, the fact that the $V$-resonance only decays through the $Wb$ channel restrict its search and makes it different for previous studies on $T$-single production with open channels $Zt$ and $ht$. However, the results of this work are also valid for a charge $2/3$ resonance, provided it decays dominantly through $T\to Wb$. To distinguish between them would require a precise determination of the $b$-charge. In any case, it is worth to remark that our model predicts Right $V$ couplings which could be differentiated from other models with Left couplings through $W$ polarisation observables.

\section*{Acknowledgments}
We thank Ricardo Piegaia for many useful discussions and suggestions, we also thank Juan Guerrero for collaboration in the early stages of this work. We are partly supported by FONCYT under the contract PICT-2010-1737 and CONICET under the contract PIP 114220100100319.

\appendix

\section{Fermion mass matrices and couplings}
\label{ap_diag}
In this appendix we show the mass matrices of the fermions and give a
numerical example to illustrate the  expected spectrum and size of
couplings. Each multiplet of composite fermions and gauge bosons has a
composite mass $m_{\phi^{cp}}$ of order TeV generated by the strong
dynamics, {\it ex}: $m_2$ and $m_b$ are respectively the masses of
$q^{cp}_2$ and $b^{cp}$. Associated to each fermionic resonance there
is also a mass mixing $\Delta_{\phi^{cp}}$. Finally there are two
composite Yukawa couplings, $y_{cp}^t$ and $y_{cp}^b$, respectively
for the top and bottom masses. We define the following basis, for
up-type fermions $T_{2/3}$: $(t^{el}, U^{cp}_1, U'^{cp}_1, U^{cp}_t,
U^{cp}_2)$, for down-type fermions $B_{-1/3}$: $(b^{el}, D^{cp}_2,
D'^{cp}_2, D^{cp}_b, D^{cp}_1)$, for $V_{-4/3}$-type fermions:
$(V'^{cp}_2, V''^{cp}_2, V'^{cp}_b)$, whereas there is one exotic
fermion $X_{5/3}$ and one $S_{-7/3}$. The corresponding $LR$ mass
matrices are:
\begin{align}
&M_T=\left(
\begin{array}{ccccc}
 0 & -\Delta_1 & 0 & 0 & -\Delta_2 \\
 0 & m_1 & 0 & y_{cp}^u\frac{v}{\sqrt{2}} & 0 \\
 0 & 0 & m_1 & y_{cp}^u\frac{v}{\sqrt{2}} & 0 \\
 -\Delta_t & y_{cp}^u\frac{v}{\sqrt{2}} & y_{cp}^u\frac{v}{\sqrt{2}} &
m_t  & 0 \\
 0 & 0 & 0 & 0 & m_2
\end{array}
\right) \ ,
\end{align}
\begin{align}
&M_B=
\left(
\begin{array}{ccccc}
0 & -\Delta_2 & 0 & 0 & -\Delta_1 \\
0 & m_2 & 0 & \sqrt{\frac{2}{3}} y_{cp}^b v & 0\\
0 & 0 & m_2 & \sqrt{\frac{1}{3}} y_{cp}^b v & 0 \\
-\Delta_d & \sqrt{\frac{2}{3}} y_{cp}^b v & \sqrt{\frac{1}{3}}
y_{cp}^b v & m_b & 0\\
0 & 0 & 0 & 0 & m_1
\end{array}
\right) \ ,
\end{align}
\begin{align}
&m_V=
\left(
\begin{array}{ccc}
m_2 & 0 & -\sqrt{\frac{1}{3}} y_{cp}^b v \\
0 & m_2 & -\sqrt{\frac{2}{3}} y_{cp}^b v \\
-\sqrt{\frac{1}{3}} y_{cp}^b v & -\sqrt{\frac{2}{3}} y_{cp}^b v & m_b
\end{array}
\right) \ ,
\end{align}
\begin{align}
& M_X=m_1 \ , \qquad M_S=m_2.
\end{align}

As an example, we show below the couplings and masses for a point of
the parameter space that solves the $A_{FB}^b$ anomaly without
spoiling the agreement with $R_b$, as well as leading to the proper
spectrum of the third generation. The input parameters are: mixing
$\sin\theta_1=0.61$, $\sin\theta_t=0.58$, $\sin\theta_2=0.045$,
$\sin\theta_b=0.8$, composite scale $M_{cp}=2$ TeV, composite Yukawa
couplings $y_{cp}^t=3, y_{cp}^b=1, g_{cp}/g_{el}=8$. The resulting
spectrum is:
\begin{eqnarray}
m_{V_i}=\{1.16,2.00,2.04\} {\rm TeV}\ ; \qquad m_S=2.00{\rm TeV} \ ;
\qquad m_X=1.58{\rm TeV} \ ; \\
m_{D_i}=\{0.0045,1.85,1.98,2.01,2.16\}{\rm TeV} \ ; \qquad
m_{U_i}=\{0.150,1.34,1.84,2.00,2.44\}{\rm TeV} \ .
\end{eqnarray}
The couplings $Wb\bar V_1$ are: $g_R\simeq 0.04$, $g_L\simeq 2\times10^{-4}$.


\end{document}